\documentclass[aps,prd,reprint,raggedbottom,floatfix,superscriptaddress,longbibliography,nofootinbib,amsmath,amssymb]{revtex4-2}
\usepackage[T1]{fontenc}
\usepackage{CJKutf8}
\makeatletter
\AddToHook{cmd/CJK@addcmap/before}{\pdfnobuiltintounicode\font@name}
\makeatother
\usepackage{microtype}
\usepackage{mathtools}
\usepackage{booktabs}
\usepackage{fontawesome5}
\usepackage{xspace,xcolor,graphicx}
\usepackage{varwidth}
\usepackage{tikz,pgfplots}
\usetikzlibrary{arrows.meta,positioning}
\pgfplotsset{compat=1.18}
\usepackage{listings}
\usepackage{hyperref}
\usepackage[nameinlink,capitalise]{cleveref}
\graphicspath{{./}}
\newcommand{\orcid}[1]{\begingroup
  \hypersetup{hidelinks}\href{https://orcid.org/#1}{\includegraphics[width=10pt]{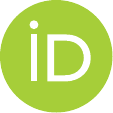}} \endgroup}
\definecolor{agentblue}{HTML}{265D85}
\definecolor{agentteal}{HTML}{278578}
\definecolor{agentgold}{HTML}{D59A32}
\definecolor{agentgray}{HTML}{F3F5F7}
\hypersetup{colorlinks=true,allcolors=blue,
 pdftitle={NuSrcAgent: An AI Agent for High-Energy Neutrino Source Searches}}

\newcommand{\system}{{\tt NuSrcAgent}\xspace}
\newcommand{\code}[1]{\texttt{#1}}
\newlength{\promptlabelwidth}
\newcommand{\promptbox}[1]{%
  \settowidth{\promptlabelwidth}{}%
  \fcolorbox{black!35}{black!14}{%
    \begin{varwidth}[t]{\dimexpr\linewidth-\promptlabelwidth-2\fboxsep-2\fboxrule-4pt\relax}%
      \small\strut\code{#1}%
    \end{varwidth}%
  }\par\nobreak\smallskip%
}
\newcommand{\status}[1]{\texttt{#1}}
\lstdefinestyle{agentcode}{basicstyle=\ttfamily\small,backgroundcolor=\color{agentgray},
 frame=single,rulecolor=\color{black!18},breaklines=true,columns=fullflexible,
 keepspaces=true,showstringspaces=false,xleftmargin=0.3em,xrightmargin=0.3em}
\begin{document}
\begin{CJK*}{UTF8}{gbsn}
\title{{\tt NuSrcAgent}: An AI Agent for High-Energy Neutrino Source Searches}

\author{Bei Zhou \orcid{0000-0003-1600-8835}\,}
\email{beizhou.email@gmail.com}
\email{beizhou@psu.edu}
\affiliation{Department of Physics, The Pennsylvania State University, University Park, Pennsylvania 16802, USA}
\affiliation{Institute for Gravitation and the Cosmos, The Pennsylvania State University, University Park, Pennsylvania 16802, USA}
\affiliation{Theory Division, Fermi National Accelerator Laboratory, Batavia, Illinois 60510, USA}
\affiliation{Kavli Institute for Cosmological Physics, University of Chicago, Chicago, Illinois 60637, USA} 

\author{Po-Wen Chang \orcid{0000-0003-1134-0652}\,}
\affiliation{National Energy Research Scientific Computing Center (NERSC), Lawrence Berkeley National Laboratory, Berkeley, California 94720, USA}

\author{Kohta Murase \orcid{0000-0002-5358-5642}\,}
\affiliation{Department of Physics, The Pennsylvania State University, University Park, Pennsylvania 16802, USA}
\affiliation{Department of Astronomy and Astrophysics, The Pennsylvania State University, University Park, Pennsylvania 16802, USA}
\affiliation{Institute for Gravitation and the Cosmos, The Pennsylvania State University, University Park, Pennsylvania 16802, USA}
\affiliation{Center for Gravitational Physics and Quantum Information, Yukawa Institute for Theoretical Physics, Kyoto, Kyoto 606-8502 Japan}
\affiliation{Department of Physics, The University of Tokyo, Bunkyo, Tokyo 133-0033, Japan}

\date{\today}

\begin{abstract}
Identifying the sources of high-energy cosmic neutrinos is crucial for understanding cosmic particle accelerators and probing new physics.
IceCube provides public data that allow researchers to search independently for neutrino sources, but these analyses require substantial technical expertise and careful scientific choices.
We present \system, the first publicly available AI agent for high-energy neutrino source searches.
Compared to public likelihood codes, \system is easier to install and use, and supports more analysis types. To start an analysis, users need only specify a source or catalog.
The agent takes natural-language requests, runs likelihood analyses on public IceCube event data, and saves the settings and calculation records so users can repeat the analyses.
We demonstrate \system through single-source and catalog stacking analyses, obtaining results consistent with IceCube results.
These examples also show its ability to search the literature, analyze non-power-law spectra using user-provided model code, and develop new analysis functions, such as signal-injection simulations.
\system is also a powerful tool for improving and testing existing analysis code, as we demonstrate by using it to refine the background probability density functions and reduce residuals and spurious signal contributions in stacking analyses.
\system makes high-energy neutrino data analysis more accessible to the broader research community.
Code and usage instructions: \href{https://github.com/beizhouphys/NuSrcAgent}{\faGithub\ {\tt github.com/beizhouphys/NuSrcAgent}}.
\end{abstract}
\maketitle

\section{Introduction}
\label{sec:introduction}

The detection of high-energy neutrinos of astrophysical origin has opened a new observational window on the Universe~\cite{Aartsen:2013bka,IceCube:2013low,IceCube:2020acn}.\footnote{
The 2026 Nobel Prize in Physics recognized Francis Halzen’s contributions to IceCube and this breakthrough~\cite{nobel2026}.
}
Identifying the sources of these neutrinos is crucial for understanding cosmic particle accelerators and probing new physics~\cite{Ackermann:2022rqc}.
Produced when accelerated ions interact with gas or radiation, these neutrinos trace hadronic activity and help identify the origins of high-energy cosmic rays~\cite{Ackermann:2019ows}.
They can also escape dense environments, which may even reveal particle acceleration hidden from electromagnetic observations~\cite{Murase:2015xka}, and candidate sources include the vicinity of supermassive black holes~\cite{Murase:2022feu} and choked jets embedded inside a massive star~\cite{Meszaros:2001ms,Ando:2005xi,Murase:2013ffa,Senno:2015tsn,Chang:2022hqj}.
Knowing the source distances and directions will establish propagation baselines for tests of new-physics scenarios~\cite{Ackermann:2022rqc}.
Evidence has emerged for neutrino emission from individual sources such as NGC~1068~\cite{IceCube:2022der} and from source populations such as X-ray-bright Seyfert galaxies~\cite{IceCube:2025gdd}.
\textit{The sources of most of the diffuse astrophysical neutrino flux remain unidentified, motivating continued searches across source populations~\cite{Murase:2016gly,IceCube:2018omy,Hooper:2018wyk,Yuan:2019ucv,IceCube:2019xiu,Creque-Sarbinowski:2021nil,Desai:2023cmq,Bouri:2024ctc,IceCube:2024tec,PugazhendhiAD:2025cjj}}. 

The IceCube Collaboration's public data releases allow researchers to investigate these questions independently~\cite{Senno:2017vtd,Esmaili:2018wnv,Hooper:2018wyk,Zhou:2021rhl,Chang:2022hqj,Desai:2023cmq,Bouri:2024ctc,PugazhendhiAD:2025cjj,Ouyang:2026abo,Mukhopadhyay:2026qlq}.
Early data releases for neutrino source searches provided the first year of data from the completed detector~\cite{ic86firstyear} and a three-year track sample~\cite{earlythreeyear}.
The subsequent 10-year release, named IceTracks-DR1, used the original Pass1 processing and supplied event data and detector response for source searches~\cite{IceCube:2021xar,dr1web}.
Most recently, the IceCube Collaboration extended the public dataset to 14 years and used Pass2 calibration and event processing for data collected since June 2010, with an updated detector response~\cite{Abbasi:2026ehs,dr2web}. This dataset is named IceTracks-DR2.
In addition, the IceCube Collaboration has released {\tt SkyLLH} code for single-source analyses using its public data, while stacking analyses of multiple sources require additional implementation~\cite{IceCube:2021mzg,IceCube:2023ihk,skyllhv26code}.

Analyzing high-energy neutrino data requires technical expertise and consistent scientific choices.
Even with public data and likelihood software, researchers must translate their neutrino emission hypotheses into a likelihood analysis, configure the fit, and evaluate the significance of the result.
Researchers must repeat much of this work to test different sources or weighting schemes, risking mistakes such as carrying over settings from an earlier analysis.
These technical demands can make source searches difficult for those who do not specialize in neutrino data analysis.
Addressing these challenges would allow more researchers to test both astrophysical source models and new-physics scenarios.
\textbf{High-energy neutrino data analysis should therefore be made more accessible to the broader community of astronomers, astrophysicists, and particle physicists}.

AI agents that use large language models (LLMs) can make neutrino data analysis more accessible by helping researchers handle these technical tasks.
They can turn researchers' natural-language requests into calculations with existing scientific tools, automate repeated operations, and help examine the results.
With appropriate instructions and training, agents can also develop additional code to perform analyses that researchers request but the existing software does not support yet.
They can save the analysis inputs and settings used to obtain each result, so researchers can check how it was calculated and repeat the analysis.
For example, {\tt HEPTAPOD} coordinates theoretical model construction, event simulation, and data analysis within a high-energy-physics workflow~\cite{Menzo:2025cim}.
{\tt FERMIACC} proposes particle-physics explanations for data and performs calculations to test their predictions~\cite{Agrawal:2026lvg}.
{\tt LeWRON} starts from an input Lagrangian and constructs model-specific code to study the electroweak phase transition and predict signals~\cite{Wang:2026jjn}.
{\tt cmbagent} infers cosmological parameters from supernova data~\cite{cmbagent}.
\textit{These capabilities motivate us to develop an AI agent for high-energy neutrino data analysis.}

In this paper, we present \system, an AI agent that carries out high-energy neutrino source searches through natural-language conversation.
Compared to public likelihood codes (e.g., {\tt SkyLLH}), \system is easy to install and use (Sec.~\ref{sec:examples}).
Users can ask a coding agent to install \system and then conduct their analyses in the same conversation.
\system also supports more analysis types, such as stacking analysis for a catalog of sources.
To start an analysis, users need only specify a source or catalog.
The agent then translates their request into an analysis using public IceCube event data and likelihood analysis code developed for \system.
It coordinates the analysis, runs the fit, and reports the fitted quantities with their statistical interpretation, so users can test and compare physical hypotheses without managing the technical procedures themselves.
It saves calculation records so users can audit and reproduce the results.
\system is also a powerful tool for improving and testing existing analysis code, as we demonstrate by using it to refine the background probability density functions and reduce residuals and spurious signal contributions in stacking analyses (Sec.~\ref{sec:background}).
\textit{It is also well instructed to write code for new analysis functions when users request analyses that the existing software does not yet support.}
\textit{Beyond these analysis functions, \system can search and examine relevant literature to identify previous studies and their analysis methods.}
Users can also ask \system directly about its capabilities, limitations, and how to use it, without initiating an analysis.

This paper is organized as follows.
Section~\ref{sec:data} provides a brief overview of the data and likelihood analysis.
Section~\ref{sec:agent} describes the workflow and user interactions with \system.
Section~\ref{sec:background} discusses the background PDFs.
Section~\ref{sec:examples} introduces the installation procedure and presents example analyses, including single-source searches with power-law and non-power-law spectra, literature searches, and a stacking analysis with signal-injection simulations, for which the agent developed new analysis code at the user's request.
Section~\ref{sec:validation} describes the validation tests.
We conclude in Sec.~\ref{sec:conclusions}.
Appendices~\ref{app:likelihood}--\ref{app:time-dependent-examples} provide further details on the likelihood formulation, statistical methods, analysis settings, calculation records, validation tests, and background PDFs, along with two concrete examples of time-dependent analysis with \system.

\begin{figure*}[!t]
\centering
\begin{tikzpicture}[
  font=\sffamily\footnotesize,
  card/.style={rounded corners=3pt,line width=.55pt,fill=white,minimum width=3.50cm,
    minimum height=1.24cm,align=center,inner sep=3pt,text=black!84},
  control/.style={card,draw=agentblue!65},
  analysis/.style={card,draw=agentteal!80,fill=agentteal!5},
  role title/.style={anchor=west,text=agentblue!75!black,
    font=\large\fontfamily{lmtt}\fontseries{b}\selectfont},
  route/.style={-{Latex[length=1.75mm,width=1.25mm]},line width=.7pt,draw=agentblue},
  computation/.style={route,draw=agentteal!90!black},
  feedback/.style={route,dashed},
  record route/.style={computation,dashed},
  edge label/.style={font=\sffamily\scriptsize,inner sep=2pt,text=agentblue!60!black}
]
  \draw[rounded corners=4pt,line width=.55pt,draw=black!30,fill=black!4]
    (0,5.10) rectangle (17.58,5.95);
  \node[role title] at (.25,5.525) {User};
  \draw[black!20,line width=.45pt] (2.36,5.28) -- (2.36,5.77);
  \node[anchor=west,text=black!84] at (2.64,5.525)
    {\textbf{Research question:} sources for analysis, literature searches, or other requests};

  \draw[rounded corners=5pt,line width=.85pt,draw=agentblue,fill=agentblue!4]
    (0,-.52) rectangle (17.58,4.53);
  \node[role title] at (.25,4.15) {\let\tt\relax\system};

  \node[control,minimum width=16.90cm,minimum height=.88cm,fill=agentblue!9]
    (interface) at (8.79,3.40)
    {\textbf{Natural-language interface}\\[3pt]A general-purpose LLM interprets the request};
  \node[analysis] (settings) at (2.09,1.64)
    {\textbf{Analysis settings}\\[3pt]Check inputs and\\record choices};
  \node[analysis] (fit) at (6.55,1.64)
    {\textbf{Likelihood fit}\\[3pt]Run existing\\likelihood code};
  \node[analysis] (review) at (11.01,1.64)
    {\textbf{Programmatic checks}\\[3pt]Check fit, settings,\\and output files};
  \node[analysis] (report) at (15.47,1.64)
    {\textbf{Analysis report}\\[3pt]Report results and\\explain warnings};
  \node[control,minimum width=7.95cm,minimum height=.88cm] (record) at (4.32,.17)
    {\textbf{Saved analysis record}\\[3pt]Settings, input checksums, and fit results};
  \node[control,minimum height=.88cm] (failed) at (11.01,.17)
    {\textbf{Report failure}\\[3pt]Explain which checks failed};

  \draw[route] (3.12,5.10) -- node[edge label,right,pos=.24]{Request and replies} (3.12,3.84);
  \draw[feedback] (8.27,3.84) -- node[edge label,right,pos=.76]{Clarification questions} (8.27,5.10);
  \draw[route] (13.12,3.84) -- node[edge label,right,pos=.76]{Progress and reports} (13.12,5.10);

  \draw[route] (1.30,2.96) -- (1.30,2.26);
  \draw[feedback] (2.84,2.26)
    -- node[edge label,right]{Ask the user for clarification} (2.84,2.96);
  \draw[computation,dashed] (fit.north)
    -- node[edge label,right]{Report progress} (6.55,2.96);

  \draw[computation] (settings.east) -- node[edge label,above]{\texttt{READY}} (fit.west);
  \draw[computation] (fit.east) -- (review.west);
  \draw[computation] (review.east) -- (report.west);
  \draw[route] (report.north) -- (15.47,2.96);
  \draw[record route] (settings.south) -- (settings.south |- record.north);
  \draw[record route] (fit.south) -- (fit.south |- record.north);
  \draw[route] (review.south) -- node[edge label,right]{Fail} (failed.north);
\end{tikzpicture}
\caption{
Architecture of \system.
}
\label{fig:architecture}
\end{figure*}

\section{Neutrino Data and Likelihood Formulation}
\label{sec:data}

This section gives a brief overview of the high-energy neutrino data and likelihood analysis used by \system.
It introduces the essential concepts and quantities needed to understand the rest of the main text.
For more details, see appendices~\ref{app:likelihood} and \ref{app:statistics}.

\system searches for neutrino emission from user-specified sources using public IceCube track events.
By default, it uses the 14-year IceTracks-DR2 sample, which covers 2008--2022 and contains 1\,643\,355 selected events~\cite{Abbasi:2026ehs,dr2web}.
The sample contains data from the IC40, IC59, IC79, and IC86 detector phases, with 36\,900, 107\,011, 101\,956, and 1\,397\,488 events, respectively.
The numbers in the phase names indicate the number of instrumented detector strings.
Each event provides a reconstructed direction, an angular uncertainty, an arrival time, and an energy proxy (i.e., the reconstructed energy of the muon produced in a neutrino interaction).
The release also provides the uptime tables, effective area, and smearing matrix for each detector phase, which connect a neutrino spectrum to the detectable signal.
\system also supports the 2008--2018 IceTracks-DR1 data~\cite{IceCube:2021xar,dr1web} for reproducing earlier analyses.
Its event reconstructions and detector response differ from IceTracks-DR2, so selecting the first ten years of IceTracks-DR2 does not exactly reproduce an IceTracks-DR1 analysis, although the differences are small.

\system uses the standard unbinned likelihood adopted by IceCube and other collaborations~\cite{Braun:2008bg,IceCube:2010nca,ANTARES:2015moa}.
For a neutrino spectrum with $m\geq0$ shape parameters $\boldsymbol{\theta}=\left\{\theta_1,\ldots,\theta_m\right\}$, the likelihood integrated over the selected observation period is
\begin{equation}
 \begin{multlined}
 \mathcal{L}(n,\boldsymbol{\theta})=\\
 \prod_k\prod_{i=1}^{N_k}
 \left[\frac{n f_k(\boldsymbol{\theta})}{N_k}S_i^k(\boldsymbol{\theta})
 +\left(1-\frac{n f_k(\boldsymbol{\theta})}{N_k}\right)B_i^k\right],
 \end{multlined}
 \label{eq:overview_likelihood}
\end{equation}
where $k$ indexes data from different detector phases and $i$ labels one of the $N_k$ selected events from detector phase $k$.
The parameter $n$ is the total expected signal count across all detector phases and sets the normalization of the neutrino spectrum (for brevity, we use $n$ instead of the usual IceCube notation $n_s$).
The quantity $f_k(\boldsymbol{\theta})$, defined in Eq.~\eqref{eq:datasetfraction}, is the fraction of the total expected signal, $n$, from detector phase $k$, with $\sum_k f_k(\boldsymbol{\theta})=1$.
These fractions depend on the livetimes and detector acceptance for the adopted neutrino spectrum.
The set $\boldsymbol{\theta}$ can contain one or more spectral shape parameters, or be empty ($m=0$) for a fixed spectrum.
Maximizing the likelihood gives $\widehat n$ and, when fitted, $\widehat{\boldsymbol{\theta}}$.
For a fixed spectrum, \system fits only $n$.
See Appendix~\ref{app:likelihood} for more details about the likelihood.

We denote the differential neutrino flux by $\phi_\nu(E_\nu;\boldsymbol{\theta})$.
\textit{Unless otherwise stated, neutrino fluxes in this paper refer to the sum of muon-neutrino and muon-antineutrino fluxes.}

A power-law neutrino spectrum, $\phi_\nu(E_\nu;\boldsymbol{\theta})\propto E_\nu^{-s}$, is commonly assumed in high-energy neutrino source searches~\cite{Braun:2008bg,IceCube:2010nca,Zhou:2021rhl,Chang:2022hqj,IceCube:2023ihk}.
\system adopts this spectrum by default, with the spectral index $s$ as its only shape parameter, i.e., $\boldsymbol{\theta}=(s)$.
We use $s$ instead of the usual IceCube notation $\Gamma$ for the power-law spectral index to avoid confusion with the Lorentz factor.
More generally, \system can accommodate user-specified neutrino spectra of arbitrary shape.

The quantities $S_i^k(\boldsymbol{\theta})$ and $B_i^k$ in Eq.~\eqref{eq:overview_likelihood} are the signal and background probability density functions (PDFs) for event $i$ from detector phase $k$, corresponding to an event from the source(s) or the background, respectively.
The signal PDF factorizes as $\mathcal{S}_{ij}^k=S_{ij,\mathrm{spatial}}^k S_{ij,\mathrm{energy}}^k S_{ij,\mathrm{temporal}}^k$, where the factors are the signal spatial, energy, and temporal PDFs, respectively.
The background PDF factorizes as $\mathcal{B}_i^k=B_{i,\mathrm{spatial}}^k B_{i,\mathrm{energy}}^k B_{i,\mathrm{temporal}}^k$, where the factors are the background spatial, energy, and temporal PDFs, respectively.

The likelihood-ratio test statistic (TS) is defined as $\mathrm{TS}=2\ln[\mathcal{L}(\widehat n,\widehat{\boldsymbol{\theta}})/\mathcal{L}(0)]$, where the numerator is the likelihood of the best-fit signal-plus-background hypothesis and the denominator is the likelihood of the background-only hypothesis.
Its $p$-value is the probability that background alone produces a TS at least as large as the observed value.
See Appendix~\ref{app:statistics} for more details about the statistical interpretation and the conversion from fitted signal counts to fluxes.

\section{Analysis workflow}
\label{sec:agent}

This section describes the overall workflow and natural-language interface (Sec.~\ref{sec:overall_workflow}), then explains how \system checks analyses and saves records for reproducibility (Sec.~\ref{sec:analysis_checks}).
The remaining subsections describe the user's interaction with \system before (Sec.~\ref{sec:before_analysis}), during (Sec.~\ref{sec:during_analysis}), and after (Sec.\ref{sec:after_analysis}) an analysis.
Users can also ask \system directly about its capabilities, limitations, and how to use it, without initiating an analysis.

\subsection{Overall workflow and natural-language interface}
\label{sec:overall_workflow}
\label{sec:natural_language_interface}

Figure~\ref{fig:architecture} shows how \system turns a research question into an analysis and a reported result.
The user describes the sources and neutrino emission hypothesis to test.
The agent inspects the inputs and records the adopted settings before calling the likelihood code.
A separate Python function checks the fit results and output files before the agent reports the results.
Section~\ref{sec:analysis_checks} describes these checks and how users can repeat the calculation.

\system uses a general-purpose LLM as the natural-language interface between the user and the analysis tools.
Users access this interface through {\tt Codex}~\cite{openai_codex} or {\tt Claude Code}~\cite{anthropic_claude_code}.
A local set of instructions, called a skill, defines the supported analyses and scientific defaults.
The skill also specifies how the agent uses the analysis tools.

Users can submit requests in English or any other language that the LLM understands, including mixed-language requests.
For each request, the AI agent determines the intended analysis and requested outputs.
When the user's request is ambiguous, the agent asks for clarification.

\subsection{Analysis checks and reproducibility}
\label{sec:analysis_checks}

For the likelihood analyses, \system uses existing likelihood code.
For example, when a user requests the same type of fit for a different source, the agent passes the new coordinates to the same fitting code.
The AI agent converts the adopted choices into explicit command arguments for \path{run_agent_wrap.py}, which serves as the single entry point for all supported analyses.
The wrapper validates these structured inputs and saves the original user request with the provenance of each setting.
It then creates a persistent analysis task.
If the task lacks a required input or scientific choice, the wrapper returns \status{AWAITING\_CONFIRMATION} and the agent asks the user for the missing information.
The wrapper starts the fit only after the task reaches \status{READY}.

After the likelihood fit, a separate Python function reads the recorded analysis configuration and the saved fit results.
It applies fixed rules to compare the results with the settings recorded before the fit and to check that the required output files exist.
For example, if a user specifies a fixed spectral index for an IceTracks-DR2 single-source fit, the function checks that the saved result records both the fixed fit mode and the specified value of $s$.
The review function also checks that the fitted quantities are finite and within their allowed ranges.
If the review issues a warning, the agent reports the result as preliminary and explains the warning.
For a recoverable software or file-access error, the wrapper may rerun the analysis once with the same settings.
If the failure persists or requires a scientific decision, the wrapper marks the task as failed, and the agent explains the cause and what needs to be resolved before another attempt.
If proceeding requires a change to the scientific settings, the agent asks the user before running the revised analysis.

\system saves the analysis settings, input checksums, software versions, and fitted quantities.
Users can inspect these records to see how the result was obtained and repeat the fit with the same inputs, settings, and software without asking the LLM to interpret the request again.

\subsection{Before the analysis: inputs and scientific choices}
\label{sec:analysis_interface}
\label{sec:before_analysis}

\system version 1.0.0 supports $2\times3\times3$ basic types of source analysis.
The ``2'' refers to the source variability: 
(1) time-integrated analysis for steady sources over the selected observation period and (2) time-dependent analysis for transients, which uses temporal PDFs with specified or fitted light-curve parameters.
The first ``3'' refers to how the source(s) are treated:
(1) a single-source analysis searches for emission from one direction,
(2) a catalog-wide single-source analysis (hereafter \textit{catalog scan}) analyzes each source in a catalog independently and applies a trial-factor correction,
(3) a stacking analysis searches for the combined neutrino emission from a source catalog.
The second ``3'' refers to how neutrino spectral information is used:
(1) a spatial+energy analysis fits both $n$ and $s$ for a power-law neutrino spectrum, which is the default option,
(2) a spatial+energy analysis fits $n$ with a fixed neutrino spectrum,
(3) a spatial-only analysis fits $n$ without the energy PDFs.

Users can provide analysis inputs and settings in various ways, such as entering them in the conversation or pointing the agent to a text file, spreadsheet, or webpage.

{\it The minimum information required from the user is the sky coordinates or names of the sources.}
Users may provide equatorial or Galactic coordinates in degrees or radians.
For right ascension (RA), \system also accepts hour-minute-second notation (e.g., 05h 34m 32s or 05:34:32).
The agent converts the coordinates to the International Celestial Reference System (ICRS) RA and declination used by the analysis.
It saves the original coordinates together with the conversion, so users can check how the supplied positions entered the calculation.
If the user supplies only names, the AI agent can look up the coordinates and ask the user if a name identifies more than one object.
If the user provides one source, \system will perform a single-source analysis.
For a catalog, the user can request a catalog scan, stacking analysis, or both.
The agent checks that the source data are valid and identifies duplicate entries.
If the catalog contains missing or invalid coordinates, the agent reports the problem and waits for the user to correct the catalog before starting the analysis.

For the two kinds of spatial+energy analyses, if the user does not specify a spectrum, \system uses the default power-law fit.
For the fixed-spectrum case, users can supply a power law with a specified $s$ or a fully specified non-power-law shape (e.g., from theoretical modeling; see Sec.~\ref{sec:ngc1068_nonpl}).
A spatial-only analysis uses event directions only and adopts a fixed spectrum ($E_\nu^{-2.5}$ by default) to calculate detector acceptance; the fit results depend weakly on the spectrum index.
The current analyses use a common spectral shape for the selected sources.

Users may also request additional spectral parameters, as shown in Eq.~\eqref{eq:overview_likelihood}, such as two parameters for a broken power law or a power law with an exponential cutoff.
Fitting these additional parameters requires code extensions and validation.
\system is well instructed to assist the users with that development.

The spatial+energy analysis uses signal energy PDFs and detector acceptance calculated for the adopted spectrum.
For a conventional power law covered by the existing signal energy PDF tables, which span $-6.1\leq-s\leq-0.9$, it interpolates those tables.
This coverage includes interpolation margins around the default fit range $-6\leq-s\leq-1$.
For a power law outside their range or a supplied non-power-law spectrum, the agent first prepares the corresponding signal energy PDFs and detector acceptance.

For the source weighting scheme used in stacking analysis, the agent adopts uniform weighting unless the user supplies relative weighting factors or specifies another scheme.
Users can choose weighting factors proportional to, e.g., fluxes in an electromagnetic band, $1/D_L^2$, or relative neutrino fluxes predicted by a theoretical model (Sec.~\ref{sec:ngc1068_nonpl}).
For $1/D_L^2$ weighting, users can supply luminosity distances $D_L$ or redshifts $z$.
The agent converts the latter to $D_L$ using a specified cosmology.
If the requested flux is ambiguous, for example when both gamma-ray and radio flux columns are available, the agent asks the user which to use.

The user's chosen weighting scheme could assign most of the assumed neutrino emission to a few sources, even in a large catalog. 
Before fitting, the agent reports the largest normalized weighting factor ($w_{\mathrm{max},\mathrm{model}}$) and the effective source count ($N_{\mathrm{eff}}^{\mathrm{src}}$) for the selected catalog column so users can assess whether it matches their neutrino emission hypothesis. 
For $N$ sources, let $w_{j,\mathrm{model}}\geq0$ denote the normalized weighting factor of source $j$, with $\sum_{j=1}^{N}w_{j,\mathrm{model}}=1$.
These diagnostics are
\begin{align}
 w_{\mathrm{max},\mathrm{model}}&\equiv\max_j w_{j,\mathrm{model}},
 \label{eq:maximum_source_weight}\\
 N_{\mathrm{eff}}^{\mathrm{src}}&\equiv\frac{1}{\sum_{j=1}^{N}w_{j,\mathrm{model}}^2}.
 \label{eq:effective_source_count}
\end{align}
For example, if one source carries 99\% of the total weighting factor, $w_{\mathrm{max},\mathrm{model}}=0.99$ and $N_{\mathrm{eff}}^{\mathrm{src}}\simeq1.02$, even for a catalog containing, e.g., 1,000 sources. 
See Appendix~\ref{app:weighting_choices} for more details about these diagnostics and additional examples.

The agent determines which parameters to fit and which to hold fixed.
Users can also specify which events enter the fit and set the ranges of the fitted parameters.
See appendix~\ref{app:event_selection} for more details about event selection.
The agent asks the user to clarify any remaining choices that affect the scientific interpretation.
For each analysis setting, the agent records whether the user specified it, whether the agent determined it from the request or supplied information, or whether it used a documented default.
It records a setting as confirmed by the user only when the user explicitly confirms that choice. 
For more details about the command conventions, see Appendix~\ref{app:interface}.

\system supports parallel computing, and users can tell the agent how many CPU threads to use.
It uses one CPU thread as the default except for background trials, which use two because they require many repeated fits and take substantial computation time.
The agent records this computing setting separately from the scientific settings.
See Appendix~\ref{app:interface} for more details about specifying the computing resources for an analysis.

If the user requests background trials, the agent records the analysis settings and the number of trials or stopping rule before starting.
Each trial uses scrambled data and repeats the same fitting procedure as the initial fit to the observed data.
See Appendix~\ref{app:trials} for details.

\subsection{During the analysis: monitoring the calculation}
\label{sec:during_analysis}

During an analysis, \system reports progress and estimates the remaining time once timing measurements become available.
For background trials, it estimates the remaining time only when the user specifies a total number of trials.

\subsection{After the analysis: results and interpretation}
\label{sec:after_analysis}

We used automated Python tests to check that \system runs the analysis with the requested settings and reports values that match the saved fit results.
To test how the agent handles user requests, we gave it a set of predefined requests.
A separate {\tt Codex} session compared the agent's replies and tool calls with the expected behavior for each request.
For example, we requested flux weighting for a catalog with both radio and X-ray flux columns without specifying which column to use.
The separate session checked the reply to see whether the agent asked the user which column to use, and checked the tool calls to confirm that the agent waited for the user's choice before starting the analysis.
See Sec.~\ref{sec:validation} and Appendix~\ref{app:validation} for more details about these tests.

In its analysis report, \system presents the fitted parameters and states whether the significance comes from an analytic approximation or background trials.
For analytic estimates, \system explains that a more precise significance requires background trials, emphasizes this when the estimate exceeds $2\sigma$, and can run these trials upon request using the same fit settings.
When background trials are performed, \system reports the empirical $p$-value, the corresponding one-sided Gaussian significance, Monte Carlo uncertainty, and a figure showing the TS distribution.
Determining the post-trial significance of the largest TS from a catalog scan requires repeating the full scan in every trial.
For more details about the statistical interpretation, see Appendix~\ref{app:statistics}.

For each reported result, \system saves the calculation record.
Users can inspect these files without rerunning the analysis.
See Appendix~\ref{app:saved_results} for more details about the saved results and calculation records.
When the user changes a scientific setting, \system saves the new analysis separately so that both results remain available for comparison.

\section{Background PDFs}
\label{sec:background}

We provide two versions of precalculated background PDFs for the default IceTracks-DR2 data sample.
Version~1 (v1) uses the same algorithm as in the IceCube {\tt SkyLLH}\footnote{In this paper, {\tt SkyLLH} refers to version~26.0.0.} public code~\cite{IceCube:2021mzg,IceCube:2023ihk,skyllhv26code}.
It constructs the spatial PDF from a histogram in $\sin\delta$ and the energy PDF from histograms in $\log_{10}(E_{\mathrm{proxy}}/\mathrm{GeV})$ within declination bins, where $E_{\mathrm{proxy}}$ is the energy proxy (i.e., reconstructed muon energy).
The energy bin width is $\Delta\log_{10}(E_{\mathrm{proxy}}/\mathrm{GeV})=0.125$, and the $\sin\delta$ binning depends on the detector phase.
The code then smooths the energy histograms along the energy axis and evaluates both PDFs using spline interpolation.
Version~3 (v3) uses adaptive sliding windows and applies a data-driven multiplicative correction to the background energy PDF for the IC40, IC59, and IC79 detector phases.
We developed v3 and use it as the default in \system to improve the accuracy of the background modeling, particularly for stacking analyses.

\textit{We find that the v1 background energy PDFs constructed using the {SkyLLH} algorithm leave residuals between the observed and predicted event counts in energy-proxy bins, which can lead to spurious signal contributions in stacking analyses.}
Especially for large catalogs, these background residuals can accumulate across many source directions in the likelihood and produce a large TS.
To examine this effect, we analyze random catalogs of 1000 sources drawn from an isotropic sky distribution, using IceTracks-DR2 events and equal weighting.
For these random catalogs, we expect TS values close to zero, apart from statistical fluctuations.
The v1 fits instead give large positive TS values of $30$--$75$ for these random catalogs, whereas v3 significantly reduces the energy residuals and lowers the TS values to $0$--$9$ for the same random catalogs.
Residuals remain in v3 at the lowest and highest energies, where the small number of observed events makes reliable background PDFs difficult to construct.
We will keep improving the background energy PDFs in \system.

See Appendix~\ref{app:background_details} for more details about the background-PDF construction, energy residuals, and comparisons of random catalogs.

For these reasons, \textit{\system uses the v3 background PDFs by default}.
In Sec.~\ref{sec:examples}, the v3 results agree more closely than the v1 results with the IceCube internal analysis of NGC~1068 and the IceCube results for magnetars.
We will continue to improve the background PDFs.

\section{Installation and example analyses}
\label{sec:examples}

We present example user prompts and show how \system interacts with users and produces results consistent with those of IceCube likelihood analyses.
These examples also demonstrate that \system excels at literature searches and is well instructed to write code for new analysis functions at users' request, such as signal-injection simulations, which are not yet implemented in the built-in analysis code of \system version 1.0.0.

The examples below use {\tt Codex}, but users can run \system with either {\tt Codex} or {\tt Claude Code}.
The example prompts given in each subsection use simple, casual language, similar to how users might chat with a friend.
In these examples, the \textbf{``final response''} is the message \system sends to the user after processing each prompt.
The exact interaction depends on the coding agent, model, reasoning setting, available tools, and prior conversation.

Section~\ref{sec:installation} shows the simplest way to install \system.
Section~\ref{sec:ngc1068} presents time-integrated single-source analyses of NGC~1068 with a power-law spectrum (Sec.~\ref{sec:ngc1068_pl}) and demonstrates how \system runs model code from a user-provided GitHub repository to generate a non-power-law neutrino spectrum and use it in the likelihood analysis (Sec.~\ref{sec:ngc1068_nonpl}).
Section~\ref{sec:magnetars} presents a time-integrated stacking analysis on magnetars that \system identified through a literature search and for which it wrote signal-injection code to calculate flux upper limits.
\begin{figure*}[!t]
\centering
\begingroup
\makeatletter
\def\@captype{table}
\makeatother
\caption{
Comparison between results from \system and the IceCube Collaboration for power-law fits to NGC~1068 using the 14-year IceTracks-DR2 sample.
For the flux normalization of the IceCube internal analysis, we convert the published $(n,s)$ result using the public DR2 detector response.
}
\label{tab:ngc1068_pl}
\setlength{\tabcolsep}{5pt}
\begin{tabular*}{\textwidth}{@{\extracolsep{\fill}}cccccc}
\hline\hline
Analysis & $\widehat n$ & $\widehat{s}$ & \shortstack{$\widehat\phi_{\nu,0}$ ($\nu_\mu+\bar\nu_\mu$, 1000~GeV)\\$[\mathrm{GeV}^{-1}\,\mathrm{cm}^{-2}\,\mathrm{s}^{-1}]$} & TS & Local significance \\
\hline
\system v1 backgroupd PDFs & 80.1 & 3.2 & $2.87\times10^{-14}$ & 29.2 & $5.0\sigma$ (analytic estimate) \\
{\tt SkyLLH} (IceCube public code)~\cite{Abbasi:2026ehs,skyllhngc} & 80.1 & 3.2 & $2.87\times10^{-14}$ & 29.2 & $5.1\sigma$ (background trials) \\
\system v3 backgroupd PDFs & 68.4 & 3.1 & $2.32\times10^{-14}$ & 22.9 & $4.4\sigma$ (analytic estimate) \\
IceCube internal analysis tools~\cite{IceCube:2025lev,Abbasi:2026ehs} & 71.1 & 3.1 & $2.49\times10^{-14}$ & Not reported & $4.6\sigma$ (background trials) \\
\hline\hline
\end{tabular*}
\endgroup
\par\vspace{1em}
\includegraphics[width=0.92\textwidth]{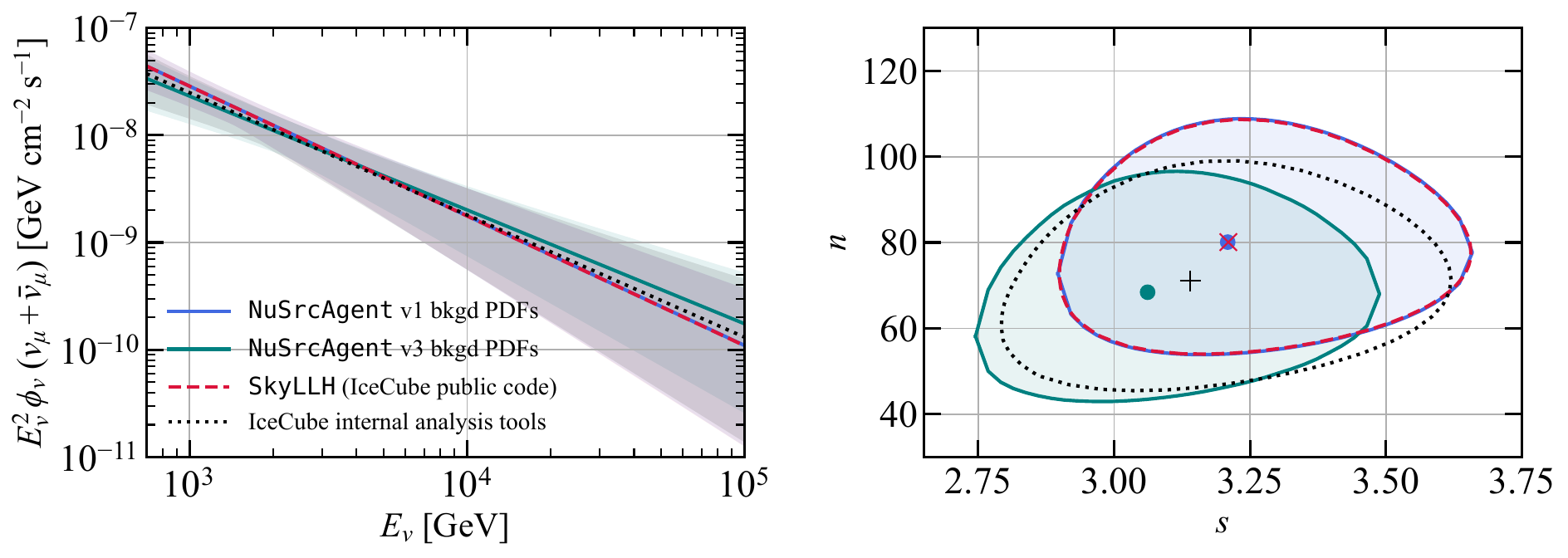}
\par\nointerlineskip
\setlength{\abovecaptionskip}{3pt}
\caption{
Comparison between results from \system and the IceCube Collaboration for power-law fits to NGC~1068 using the 14-year IceTracks-DR2 sample.
The left panel shows the flux spectra, and the right panel shows the $(n,s)$ joint confidence regions.
Spectrum envelopes and contours correspond to 68\% confidence.
For the flux spectrum and envelope of the IceCube internal analysis, we convert the published $(n,s)$ contours using the public DR2 detector response~\cite{IceCube:2025lev,Abbasi:2026ehs}.
The energy range shown in the left panel is chosen for display.
}
\label{fig:ngc1068_pl}
\end{figure*}

Compared with v1, the default v3 background PDFs give better agreement with the IceCube internal analysis of NGC~1068 (Sec.~\ref{sec:ngc1068_pl}) and with the IceCube results for magnetars (Sec.~\ref{sec:magnetars}).

\begin{figure*}[!t]
\centering
\begingroup
\makeatletter
\def\@captype{table}
\makeatother
\caption{
Results from \system for NGC~1068 using v3 background PDFs.
We fit the 14-year IceTracks-DR2 data using the neutrino spectrum predicted by the turbulent corona model~\cite{Murase:2019vdl,Murase:2026hrz}.
For the fitted turbulent corona spectrum, we use $1\leq\eta_{\rm tur}\leq1000$.
We report fitted values of $\eta_{\rm tur}$ rounded to the nearest integer.
The last row gives the power-law fit as a reference and repeats the \system v3 result in the third row of Table~\ref{tab:ngc1068_pl}.
}
\label{tab:ngc1068_km}
\setlength{\tabcolsep}{8pt}
\begin{tabular*}{\textwidth}{@{\extracolsep{\fill}}ccccc}
\hline\hline
Spectrum & $\widehat n$ & Spectral parameter & \shortstack{$\widehat\phi_{\nu,0}$ ($\nu_\mu+\bar\nu_\mu$, 1000~GeV)\\$[\mathrm{GeV}^{-1}\,\mathrm{cm}^{-2}\,\mathrm{s}^{-1}]$} & TS \\
\hline
Turbulent corona model (Model A) & 39.8 & $\eta_{\rm tur}=10$ (fixed) & $7.70\times10^{-16}$ & 16.6 \\
Turbulent corona model (Model A) & 64.2 & $\widehat\eta_{\rm tur}=173$ & $2.33\times10^{-14}$ & 21.1 \\
Turbulent corona model (Model C) & 48.4 & $\eta_{\rm tur}=15$ (fixed) & $3.01\times10^{-15}$ & 18.9 \\
Turbulent corona model (Model C) & 60.1 & $\widehat\eta_{\rm tur}=47$ & $1.47\times10^{-14}$ & 21.1 \\
Power-law spectrum & 68.4 & $\widehat{s}=3.1$ & $2.32\times10^{-14}$ & 22.9 \\
\hline\hline
\end{tabular*}
\endgroup
\par\vspace{1em}
\includegraphics[width=\textwidth]{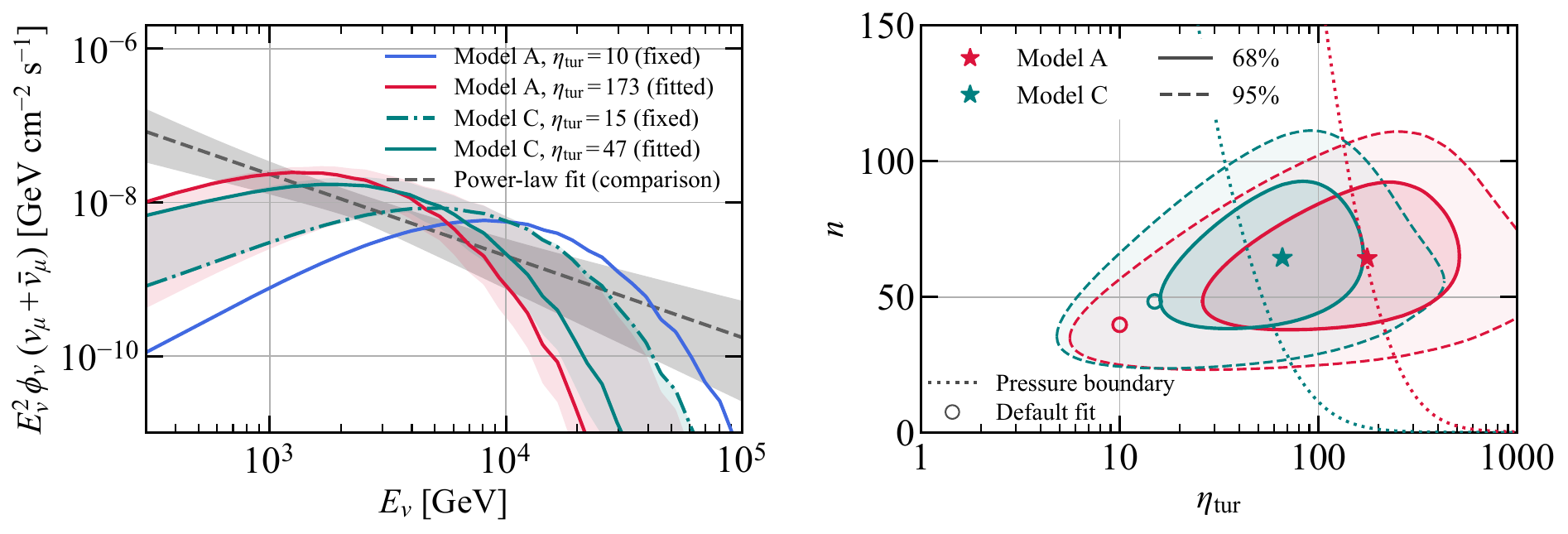}
\par\nointerlineskip
\setlength{\abovecaptionskip}{3pt}
\caption{
Results from \system for NGC~1068 using the 14-year IceTracks-DR2 data, v3 background PDFs, and a user-provided non-power-law neutrino spectrum predicted by the magnetically powered turbulent corona model~\cite{Murase:2019vdl,Murase:2026hrz}.
The left panel shows spectra for each model with $\eta_{\rm tur}$ fixed at its default value or fitted, with the flux normalization fitted in all four cases.
For the fitted turbulent corona spectrum, we use $1\leq\eta_{\rm tur}\leq1000$.
The bands for Models A and C span the fluxes predicted by their nominal 68\% joint $(n,\eta_{\rm tur})$ regions with the pressure bound ($P_{\mathrm{CR}}/P_{\mathrm{th}}=0.5$) applied.
For comparison, we show the best-fit power-law spectrum and its 68\% envelope from Fig.~\ref{fig:ngc1068_pl}.
The right panel shows the nominal 68\% and 95\% joint confidence contours and the best-fit points with free $\eta_{\rm tur}$ for Models A and C with the pressure bound removed.
Points above each model's pressure boundary have $P_{\mathrm{CR}}/P_{\mathrm{th}}>0.5$ and lie outside the parameter range used for the fits in Table~\ref{tab:ngc1068_km}.
}
\label{fig:ngc1068_km}
\end{figure*}

\subsection{Installation}
\label{sec:installation}

The easiest way to install \system is to give its GitHub repository URL to {\tt Codex} or {\tt Claude Code} and ask it to install the software for the user.
For example, users can send the following prompt.

\noindent\promptbox{Install NuSrcAgent from
{\scriptsize\url{https://github.com/beizhouphys/NuSrcAgent}}}

The coding agent will read the installation instructions in the repository, clone the code, run the installation script, download the required analysis inputs, and verify the installed files.
After installation, users can continue in the same conversation with prompts like those in the next two subsections.

\subsection{Time-integrated single-source analyses:\\ NGC 1068}
\label{sec:ngc1068}

\subsubsection{Power-law spectrum and comparison with IceCube results}
\label{sec:ngc1068_pl}

Table~\ref{tab:ngc1068_pl} and Fig.~\ref{fig:ngc1068_pl} show the comparison of power-law fits to NGC~1068 from \system and the IceCube Collaboration.
For NGC~1068, the IceCube Collaboration has reported results obtained using both the public {\tt SkyLLH} code and its internal analysis tools~\cite{IceCube:2025lev,skyllhngc} using the same IceTracks-DR2 event sample~\cite{Abbasi:2026ehs,dr2web}.
For this comparison, \system performs two analyses, one using v1 background PDFs and the other using v3 background PDFs.

We first fit the power-law spectrum, varying $n$ and $s$.
We use the public DR2 detector response and the same signal PDF definitions as in {\tt SkyLLH}~\cite{Abbasi:2026ehs,skyllhv26code}.
With v1 background PDFs, \system reproduces the {\tt SkyLLH} result, as shown in Table~\ref{tab:ngc1068_pl} and Fig.~\ref{fig:ngc1068_pl}.

Compared with v1 background PDFs, v3 background PDFs give a fitted signal count ($\hat{n}$), significance estimate, and flux measurement closer to the IceCube internal analysis results, as shown in Table~\ref{tab:ngc1068_pl} and Fig.~\ref{fig:ngc1068_pl}.
These comparisons support the choice of v3 background PDFs as the default in \system.
The remaining differences between our v3 results and the IceCube internal analysis stem from differences in signal energy PDF construction and significance calculation.
Our signal energy PDF, $P(E_{\mathrm{proxy}}\mid\delta_{\mathrm{src}},s)$, is conditional on the source declination $\delta_{\mathrm{src}}$, whereas IceCube constructs $P(E_{\mathrm{proxy}}\mid\delta_{\mathrm{rec}},s)$ from internal simulations at the reconstructed event declination $\delta_{\mathrm{rec}}$~\cite{Abbasi:2026ehs}.
  For the significance, we use an analytic estimate, whereas IceCube uses background trials.

Users can obtain the results shown in Table~\ref{tab:ngc1068_pl} and Fig.~\ref{fig:ngc1068_pl} with the following prompts.

\noindent

\setlength{\fboxsep}{2pt}
\setlength{\leftmargini}{1.3em}
\begin{enumerate}

\item \promptbox{Analyze NGC 1068}
\textbf{Explanation:} \system looks up the coordinate of NGC 1068 and fits $n$ and $s$ for a power-law spectrum using its default 14-year IceTracks-DR2 sample and v3 background PDFs.
The final response includes a results table with the best-fit $n$ and $s$, TS, and a significance estimate, together with saved numerical fit records and diagnostics.

\item \promptbox{Repeat with v1 Bpdfs}
\textbf{Explanation:} \system repeats the fit using v1 background PDFs.
The final response is similar to the above.

\item \promptbox{Measure the flux for both fits}
\textbf{Explanation:} \system measures the flux for each fit with the profile likelihood and obtains the 68\% joint $(n,s)$ regions.
The final response includes the flux measurements, plots of spectra with their uncertainty bands, and the confidence contours.

\item \promptbox{Compare these NGC 1068 fits with the SkyLLH result.
Are there other IceCube published NGC 1068 results?}
\textbf{Explanation:} 
\system finds the NGC~1068 result in the public {\tt SkyLLH} documentation and searches the literature for other IceCube NGC~1068 results. 
The final response includes a table comparing the above v1 and v3 results with the {\tt SkyLLH} result, together with a list of other published NGC~1068 results from IceCube with their event samples, analysis settings, and other relevant information. 
\system also points out that the IceCube internal analysis reported alongside the {\tt SkyLLH} result in the IceTracks-DR2 paper~\cite{Abbasi:2026ehs} is the most directly comparable because it uses the same event sample.
The user then selects it for the subsequent comparison.

\item \promptbox{Compare results from v1/3, skyllh, the icecube internal analysis results in a table and plots. One plot for the spectrum+uncertainty; one for the parameter contours.}
\textbf{Explanation:} \system compiles the fitted and reference results, converts the flux quantities to common units, and saves the numerical values used for the comparison.
The final response includes the requested outputs and links to the saved numerical values.
\end{enumerate}

\subsubsection{Non-power-law spectrum from the magnetically powered turbulent corona model}
\label{sec:ngc1068_nonpl}

To show how \system works with a non-power-law spectrum provided by the user, we use the neutrino spectrum predicted by the turbulent corona model~\cite{Murase:2019vdl,Murase:2022dog,Murase:2026hrz,Carpio:2026xkf}.
In this model, $\eta_{\mathrm{tur}}$ is regarded as an effective parameter controlling the efficiency of particle acceleration~\cite{Murase:2026hrz}.

Table~\ref{tab:ngc1068_km} and Fig.~\ref{fig:ngc1068_km} show the results for NGC~1068 from two fits for each model using v3 background PDFs.
For both Models A and C, we use $\alpha=0.1$, $\log_{10}[L_X/(\mathrm{erg\,s^{-1}})]=43.602$, and a luminosity distance of 10~Mpc~\cite{Murase:2019vdl,Murase:2026hrz,kmcoronacode}\footnote{For Model A, we use $q=5/3$, $\beta=1$, $R=3R_{\mathrm{S}}$, and $\log_{10}(M_{\mathrm{BH}}/M_\odot)=6.78$.
Here $R_{\mathrm{S}}$ is the Schwarzschild radius.}
\footnote{For Model C, we use $q=2$, $\beta=0.01$, $R=10R_{\mathrm{S}}$, and $\log_{10}(M_{\mathrm{BH}}/M_\odot)=7$.}.
We fix the proton injection energy to 1~TeV in all fits.
For the fits reported in Table~\ref{tab:ngc1068_km}, we constrain the flux normalization so that $P_{\mathrm{CR}}/P_{\mathrm{th}}\leq0.50$.
For each model, we first fix $\eta_{\mathrm{tur}}$ at its default value, 10 for Model A and 15 for Model C, and fit only the flux normalization.
In the second fit, we vary $n$ and $\log_{10}\eta_{\mathrm{tur}}$ with $1\leq\eta_{\mathrm{tur}}\leq1000$.
At their default $\eta_{\mathrm{tur}}$ values, both models give acceptable fits, as their best-fit points lie within the nominal 95\% joint confidence regions (about $2\sigma$) in Fig.~\ref{fig:ngc1068_km}.
For the turbulent corona model, \system automatically calculates the signal energy PDFs and detector acceptance for each value of $\eta_{\mathrm{tur}}$ and uses them to evaluate the likelihood, because the pre-calculated quantities are available only for power-law spectra.

Users can obtain the above results for Model A with the following prompts.

\begingroup
\noindent

\setlength{\fboxsep}{2pt}
\setlength{\leftmargini}{1.3em}
\begin{enumerate}
\addtolength{\linewidth}{-8pt}

\item \promptbox{Can you fit NGC 1068 with spectrum from Murase's turbulent corona model,
{\scriptsize\url{https://github.com/pegasuskmurase/ModelTemplates/tree/main/AGNcoronaMM_MKM20model/Distribution}}?
Try to run the github code and compute the spectrum.}
\textbf{Explanation:} \system checks the package instructions on GitHub and finds an example configuration.
The model parameters determine the spectrum used in the fit, so the final response explains how to run the generator and asks the user which parameters to use.
\par\nobreak\smallskip
\textbf{Agent question:}\enspace Do you have a set of parameters from Murase, or should I use the example in the package?

\item \promptbox{Use the example parameters in the package.}
\textbf{Explanation:} The GitHub repository contains separate macOS builds for Apple silicon ({\tt arm64}) and Intel ({\tt x86\_64}).
\system downloads the Apple silicon build for the user's computer and runs the example calculation.
The example parameter set uses Model A with $\eta_{\mathrm{tur}}=10$~\cite{Murase:2026hrz}.
\system extracts the all-flavor neutrino spectrum above $0.3$ GeV from {\tt FinalFlux.dat} and converts the energy and differential flux to GeV units.
Using the default v3 background PDFs, \system computes the signal energy PDF and detector acceptance for this spectrum and fits $n$ to the 14-year IceTracks-DR2 data.
The final response identifies the result as a package-example calculation and reports the best-fit $n$ and TS.
It links to the generated spectrum and fit records.

\item \promptbox{Now also fit eta.}
\textbf{Explanation:} \system plans to fit $\eta_{\mathrm{tur}}$ together with $n$, keeping the other model inputs fixed and generating spectra at different $\eta_{\mathrm{tur}}$ values.
Before generating these spectra, \system asks the user in its final response for the minimum and maximum $\eta_{\mathrm{tur}}$ values to test.
\par\nobreak\smallskip
\textbf{Agent question:}\enspace What range should I use for $\eta_{\mathrm{tur}}$?

\item \promptbox{1--1000. Also please fit in log10 eta instead linear.}
\textbf{Explanation:} 
\system generates spectra across the specified $\eta_{\mathrm{tur}}$ range and uses the public detector response to compute the signal energy PDF and detector acceptance for each spectrum, and then performs the likelihood analysis.
The final response gives the best fit parameters, TS, and links to the saved fit records.

\begin{table*}[!t]
\centering
\caption{
Comparison between results from \system and those reported in the IceCube Ph.D. thesis~\cite{ghadimi2025} for an equal-weighting stacking analysis of 32 magnetars assuming a power-law neutrino spectrum.
\system estimates the pre-trial $p$-values analytically, while the thesis reports $p$-values calculated from a fit to the TS distribution of background trials.
The flux upper limits apply to the total $\nu_\mu+\bar\nu_\mu$ flux from the catalog for an $E_\nu^{-2}$ spectrum and use the one-sided 90\% Neyman construction, with $n$ and $s$ fitted in each trial.
The quoted upper-limit uncertainties are Monte Carlo standard errors.
}
\label{tab:magnetar_comparison}
\setlength{\tabcolsep}{8pt}
\begin{tabular*}{\textwidth}{@{\extracolsep{\fill}}cccccc}
\hline\hline  
Analysis & $\widehat n$ & $\widehat{s}$ & \shortstack{$E_\nu^2\phi_{\nu,\mathrm{UL}}^{90\%}$ ($\nu_\mu+\bar\nu_\mu$)\\$[\mathrm{GeV\,cm^{-2}\,s^{-1}}]$} & TS & \shortstack{Pre-trial\\$p$-value} \\
\hline
IceCube Ph.D. thesis~\cite{ghadimi2025} & 53.3 & 3.2 & $1.98\times10^{-9}$ & 1.9 & 0.13 \\
\system v3 & 37.8 & 3.2 & $(1.94\pm0.12)\times10^{-9}$ & 0.9 & 0.28 \\
\system v1 & 78.6 & 3.6 & $(3.12\pm0.14)\times10^{-9}$ & 3.8 & 0.08 \\
\hline\hline
\end{tabular*}
\end{table*}

\item \promptbox{Can you do the flux measurement for this non power law spectrum in the similar way as you do for the default power law spectrum?}

\textbf{Explanation:} \system follows the flux-measurement procedure used for the default power-law fit (e.g., in Sec.~\ref{sec:ngc1068_pl}) and converts the $\widehat{n}$ to $\nu_\mu+\bar\nu_\mu$ differential fluxes.
For the fit with free $\eta_{\mathrm{tur}}$, it obtains 68\% and 95\% joint $(n,\eta_{\mathrm{tur}})$ confidence regions from the profile likelihood and uses the 68\% region to calculate a flux uncertainty band.
The final response includes the flux measurements with uncertainties, confidence regions, and links to the profile-likelihood results.

\item \promptbox{Compare/show the two turbulent corona fits with the v3 power-law fit in a table and plots. One plot for spectrum; one plot for the parameter contour.}
\textbf{Explanation:} \system retrieves the saved power-law fit resutls with the default v3 background pdfs from Sec.~\ref{sec:ngc1068_pl} and puts the fluxes in common units.
It creates a table with the spectrum, best-fit $n$, spectral parameter, flux at 1~TeV, and TS for all three fits, marking $\eta_{\mathrm{tur}}=10$ as fixed.
In one figure, it plots $E_\nu^2\phi_\nu$ with 68\% uncertainty bands.
In another figure, it plots the 68\% and 95\% joint $(n,\eta_{\mathrm{tur}})$ contours, marks the best fit with a star, and uses a logarithmic $\eta_{\mathrm{tur}}$ axis from $1$ to $1000$.
\end{enumerate}
\endgroup

\subsection{Time-integrated stacking analysis: magnetars}
\label{sec:magnetars}

After surveying the literature with \system, we select the IceCube time-integrated stacking analysis on magnetar~\cite{IceCube:2023ugc,ghadimi2025} for comparison because its data sample is closest to the public IceTracks-DR2 used by \system.
The IceCube analysis uses the internal PSTracks~V4.2 sample of muon tracks collected from April 2008 to May 2022, covering the same period as the public IceTracks-DR2 sample.
Both data samples consist of Pass2-reprocessed data~\cite{ghadimi2025,Abbasi:2026ehs}, so we expect them to be very similar.

For the IceCube magnetar analysis, the 2023 conference proceedings present sensitivity and discovery potential~\cite{IceCube:2023ugc}, while the 2025 IceCube Ph.D. thesis reports results from analyzing the data~\cite{ghadimi2025}.
The thesis provides source positions and selection criteria, weighting schemes, fitted signal counts and spectral indices, TS values, significances, and 90\% confidence-level flux upper limits.
To reproduce the thesis analysis, \system automatically replaces its default Rayleigh signal spatial PDF, also used by {\tt SkyLLH}, with the thesis's Gaussian signal spatial PDF~\cite{ghadimi2025,skyllhv26code}.

Table~\ref{tab:magnetar_comparison} compares the equal-weighting stacking results from \system using v1 and v3 background PDFs with those reported in the thesis for the same 32 magnetars.
All the analyses fit the total signal count $n$ and a common spectral index $s$.

With v3 background PDFs, \system reproduces the published spectral index to its quoted precision and gives a flux upper limit consistent with the thesis within the Monte Carlo uncertainty.
It also obtains $\widehat n$ and TS values closer to the thesis results than with v1 background PDFs.
These results support our choice of v3 background PDFs as the default in \system.
The remaining differences between \system v3 results and the IceCube internal analysis reported in the thesis are likely due to differences in the signal energy PDF construction, as in the NGC~1068 comparison in the previous subsection.

We first asked \system for approximate analytic flux upper limits for the stacking analysis on magnetars.
After it explained how the thesis used background and signal-injection trials to obtain its 90\% Neyman upper limits, we asked the agent to implement that construction, which is not yet available in its built-in analysis code.
The agent wrote and executed the required code, achieving the agreement with the thesis described above.
\textit{This example demonstrates that \system is well instructed to implement new analysis functions at users' request.}

\begingroup

The literature search and analysis above are based mainly on the following prompts.

\setlength{\fboxsep}{2pt}
\setlength{\leftmargini}{1.3em}
\begin{enumerate}
\addtolength{\linewidth}{-8pt}

\item \promptbox{Is there any literature on time-integrated stacking analyses using the IceTracks-DR2 data?
Only look up those from the IceCube Collaboration.}
\textbf{Explanation:} \system searches for stacking-analysis results from the IceCube Collaboration.
It cannot confirm that any published stacking fit used the same IceTracks-DR2 data.
The final response lists the relevant studies and identifies the IceCube magnetar search~\cite{IceCube:2023ugc,ghadimi2025,Abbasi:2026ehs} as the closest comparison because its internal PSTracks~V4.2 sample and public IceTracks-DR2 both consist of Pass2-reprocessed data and cover the same period.

\item \promptbox{Can you match NuSrcAgent's analysis to the IceCube magnetar analysis in the thesis as closely as possible?}
\textbf{Explanation:} After the user selects the magnetar search, \system examines the proceedings and thesis it found in detail~\cite{IceCube:2023ugc,ghadimi2025}.
The proceedings report sensitivity and discovery potential, while the thesis provides the source positions and analysis settings needed to match the fit.
The final response compares the event samples, identifies settings \system can match, and explains the remaining differences.

\item \promptbox{Now try to reproduce the thesis's stacking analysis. Use the same magnetars with equal weighting and match the thesis settings as closely as possible.}
\textbf{Explanation:} \system identifies the 32 magnetars used in the thesis's analysis and reads their positions from its catalog table in the thesis.
The table labels the coordinates J2000 without specifying ICRS or FK5, so the final response asks the user how to interpret them before running the fit.
\par\nobreak\smallskip
\textbf{Agent question:}\enspace Should I use the tabulated RA and declination as approximate ICRS coordinates?

\item \promptbox{Yes. And try both v3 and v1 Bpdfs.}
\textbf{Explanation:} To match the thesis, \system automatically switches its signal spatial PDF from the default Rayleigh form to a Gaussian.
It then fits the same 32 magnetars with equal weighting.
The final response includes the fitted $n$ and $s$, TS, analytic pre-trial $p$-value estimates, and links to the fit records.

\item \promptbox{Calculate the 90\% flux upper limits for the v3 and v1 fits. Do it analytically first, without running simulations.}
\textbf{Explanation:} \system calculates one-sided 90\% flux upper limits for a fixed $E_\nu^{-2}$ spectrum (same as the thesis) using the profile-likelihood.
The agent reports the limits and notes to the user that the thesis obtains its one-sided 90\% Neyman upper limits using background and signal-injection trials, not an analytic profile-likelihood approximation.

\item \promptbox{Can you implement the thesis's signal-injection method to calculate the 90\% flux upper limits the same way? Do a short test run first.}
\textbf{Explanation:} \system writes and checks the signal-injection code for the one-sided Neyman construction.
It runs and times small batches of background trials and signal-injection trials with the v3 background PDFs and repeats both batches with the v1 PDFs.
With two CPU threads, the agent estimates about 40 minutes total for 100 trials.
The final response asks the user how much Monte Carlo uncertainty is acceptable for the upper limits.
\par\nobreak\smallskip
\textbf{Agent question:}\enspace What Monte Carlo precision do you want for the flux upper limits?

\item \promptbox{10\% Monte Carlo uncertainty on the flux UL.}
\textbf{Explanation:} Using the short test run and the user's precision target, \system chooses an initial budget of 200 background trials and about 800 signal-injection trials for each background PDF version (default v3 and v1).
It estimates a total runtime of about 6--7 hours for these trials with two CPU threads.
It injects signals with an $E_\nu^{-2}$ spectrum into scrambled data at several flux levels and refits $n$ and $s$ in every trial.
It estimates the flux at which 90\% of signal trials exceed the larger of the observed TS and the median TS from background trials.
The final response reports the two flux upper limits, their Monte Carlo uncertainties, and links to the trial records.

\item \promptbox{Put the v1, v3, and thesis results in one table}
\textbf{Explanation:} \system expresses the flux upper limits in common units and uses the thesis's published $p$-value for the reference row.
The final response includes the best-fit $n$ and $s$, TS, pre-trial $p$-values, and the 90\% flux upper limits with Monte Carlo uncertainties.
\end{enumerate}
\endgroup

\section{Validation}
\label{sec:validation}

This section describes how we validate \system using software tests and selected conversations.
These tests are separate from the per-analysis checks in Sec.~\ref{sec:analysis_checks}, which the agent uses to decide whether to report each analysis result as valid.
See Appendix~\ref{app:validation} for more details.

In the automated Python tests, we check that \system runs the likelihood analysis with the requested settings and that the reported values agree with the saved fit results.
We also check that fixed-spectrum 95\% upper limits satisfy $2\Delta\ln\mathcal{L}=2.71$.
These checks passed.

We use selected conversations to check whether the AI agent understands the user's request, asks the user to clarify scientific choices, and explains the results correctly.
For example, we requested flux weighting for a catalog with both radio and X-ray flux columns without specifying a column.
We checked that the agent asked the user which column to use and waited for an answer before running the analysis.
We also review these conversations to find cases where the agent asks the same question more than once or gives an unclear response.

\section{Conclusions}
\label{sec:conclusions}
\label{sec:discussion}

Identifying the sources of high-energy cosmic neutrinos is essential for understanding cosmic particle accelerators and using known baselines (i.e., distances and directions) to test new physics.
Public IceCube data and likelihood software allow researchers to search independently for neutrino sources. However, translating a neutrino emission hypothesis into a likelihood analysis still requires substantial technical expertise and careful scientific choices. 
Researchers must also ensure that the analysis settings remain appropriate and consistent when testing different sources or weighting schemes, which can be challenging for those who do not specialize in neutrino data analysis.
An AI agent can help researchers carry out these tasks through natural-language conversation, provided that it uses consistent scientific settings throughout the calculation.

In this paper, we present \system, an AI agent for high-energy neutrino source searches using IceCube public data.
Users describe the sources and neutrino emission hypothesis they want to test, and the agent determines the analysis settings, runs the likelihood calculation, and reports the fitted quantities and their statistical interpretation.
It records how scientific settings were chosen and saves calculation records so users can inspect how a result was obtained, repeat the analysis, and compare different hypotheses.
It is also well instructed to write code for new analysis functions when users request analyses that the existing software does not yet support.
In addition, \system can search and examine relevant literature to assist users with their research.

We demonstrate \system through single-source analyses of NGC~1068 and a stacking analysis of magnetars, obtaining results consistent with the corresponding IceCube results.
We also present example user prompts and show how \system processes these requests and reports the results to users.
Beyond the basic function of doing likelihood analyses, these examples also demonstrate several additional capabilities of \system:
(1) search and examine relevant literature; 
(2) run theoretical model code from a user-provided GitHub repository to generate non-power-law neutrino spectra and incorporate them into likelihood fits; 
(3) write and run new signal-injection code at the user's request to calculate flux upper limits, demonstrating its ability to develop analysis functions not yet supported by the built-in software.

We have also performed validation tests to verify that \system correctly interprets user requests, carries out the specified likelihood analyses, and accurately reports the calculated results.
These tests include software checks and selected multilingual conversations to evaluate how the agent handles ambiguous scientific choices and explains its results. 

By enabling high-energy neutrino source searches through natural-language conversation, \textit{\system can make the high-energy neutrino data analysis more accessible to the broader research community.}
We will continue to develop and improve \system, such as additional analysis functions, particularly those requested by the research community.

\begin{acknowledgments}

We thank Isaac R. Wang~\cite{Wang:2026jjn} for helpful discussions on developing AI agents during the early stages of this project and for commenting on the manuscript.
We thank Wahid Bhimji for feedback and support with the computational resources.
We thank the IceCube Collaboration for making the event data and detector responses, and {\tt SkyLLH} code publicly available.
We used these resources to develop our code for catalog scans, stacking analyses, and other analyses.
Some of the code was developed with assistance from {\tt Codex}.
B.\,Z. was supported by NASA grants No.~80NSSC20K0775 and No.~80NSSC20K0925 and by the Institute for Gravitation and the Cosmos at The Pennsylvania State University.
While at Fermilab, B.\,Z. was supported by Fermi Forward Discovery Group, LLC under Contract No. 89243024CSC000002 with the U.S. Department of Energy, Office of Science, Office of High Energy Physics. Part of this work used resources of the National Energy Research Scientific Computing Center (NERSC), a Department of Energy User Facility.
The work of K.M. was supported by the NSF Grant No.~AST-2308021. K.M. also acknowledges Mamoru Yanagisawa for his generous donation and continuous support.
\end{acknowledgments}

\onecolumngrid
\clearpage
\appendix
\section{Likelihood formulation and probability density functions}
\label{app:likelihood}

This appendix details the likelihood analysis briefly discussed in Sec.~\ref{sec:data}.
We present the likelihood and test statistic (Appendix~\ref{app:likelihood_definition}), the signal PDFs and detector response (Appendix~\ref{app:signal_model}), and the background PDFs (Appendix~\ref{app:pdfs}).

\subsection{Likelihood and test statistic}
\label{app:likelihood_definition}

\system uses the standard unbinned likelihood adopted by IceCube and other collaborations~\cite{Braun:2008bg,IceCube:2010nca,ANTARES:2015moa}.
For a neutrino spectrum with $m\geq0$ shape parameters $\boldsymbol{\theta}=(\theta_1,\ldots,\theta_m)$, the likelihood integrated over the selected observation period is
\begin{equation}
 \begin{multlined}
 \mathcal{L}(n,\boldsymbol{\theta})=
 \prod_k\prod_{i=1}^{N_k}
 \left[\frac{n f_k(\boldsymbol{\theta})}{N_k}S_i^k(\boldsymbol{\theta})
 +\left(1-\frac{n f_k(\boldsymbol{\theta})}{N_k}\right)B_i^k\right].
 \end{multlined}
 \label{eq:likelihood}
\end{equation}
where $k$ indexes data from different detector phases and $i$ labels one of the $N_k$ selected events from detector phase $k$.
The parameter $n$ is the total expected signal count across all detector phases and sets the normalization of the neutrino spectrum (for brevity, we use $n$ instead of the usual IceCube notation $n_s$).
The quantity $f_k(\boldsymbol{\theta})$, defined in Eq.~\eqref{eq:datasetfraction}, is the fraction of the total expected signal from detector phase $k$, with $\sum_k f_k(\boldsymbol{\theta})=1$.
These fractions depend on the livetimes and detector acceptance for the adopted neutrino spectrum.
Thus, $n f_k(\boldsymbol{\theta})/N_k$ is the signal fraction in the data from the detector phase $k$ and $1-n f_k(\boldsymbol{\theta})/N_k$ is the background fraction.
The quantities $S_i^k(\boldsymbol{\theta})$ (Appendix~\ref{app:signal_model}) and $B_i^k$ (Appendix~\ref{app:pdfs}) are the signal and background probability density functions (PDFs) for event $i$ from detector phase $k$, corresponding to an event from the source(s) or the background, respectively.

\system maximizes the log-likelihood ratio,
\begin{equation}
 \ln\frac{\mathcal{L}(n,\boldsymbol{\theta})}{\mathcal{L}(0)}=
 \sum_k\sum_{i=1}^{N_k}
 \ln\!\left[1+\frac{n f_k(\boldsymbol{\theta})}{N_k}
 \left(\frac{S_i^k(\boldsymbol{\theta})}{B_i^k}-1\right)\right],
 \label{eq:logratio}
\end{equation}
where $\mathcal{L}(0)$ is the likelihood under the background-only hypothesis. 
The resulting likelihood-ratio test statistic (TS) is
\begin{equation}
 \mathrm{TS}\equiv2\left[\ln\mathcal{L}(\widehat n,\widehat{\boldsymbol{\theta}})
 -\ln\mathcal{L}(0)\right].
 \label{eq:ts}
\end{equation}
Here $\widehat n$ and $\widehat{\boldsymbol{\theta}}$ maximize Eq.~\eqref{eq:logratio} with respect to $n$ and the spectral parameters.

\system adopts a power-law neutrino spectrum by default.
For source $j$, the differential neutrino flux is
\begin{equation}
 \phi_{\nu,j}(E_\nu)=\phi_{\nu,0,j}\left(\frac{E_\nu}{E_0}\right)^{-s}.
 \label{eq:spectrum}
\end{equation}
Here $\phi_{\nu,0,j}$ is the differential flux normalization at the fixed reference energy $E_0$.

The default fit has one spectral shape parameter, $s$, and varies both $n$ and $s$, giving $\boldsymbol{\theta}=\left\{s\right\}$.
We use $s$ instead of the usual IceCube notation $\Gamma$ for the power-law spectral index to avoid confusion with the Lorentz factor.
A fixed spectral index or a fully specified non-power-law shape gives $m=0$.
For a fixed spectrum, \system fits $n$ while holding the spectral shape fixed.

\subsection{Signal PDFs and detector response}
\label{app:signal_model}

The signal PDF factorizes into spatial, energy, and temporal PDFs~\cite{Chang:2022hqj,IceCube:2023ihk},
\begin{equation}
 \mathcal{S}_{ij}^k=
 S_{ij,\mathrm{spatial}}^k(\psi_{ij}^k,\sigma_i^k)\,
 S_{ij,\mathrm{energy}}^k(E_{\mathrm{proxy},i}^k\mid\delta_j,\boldsymbol{\theta})\,
 S_{ij,\mathrm{temporal}}^k(t_i^k),
 \label{eq:signal_factors}
\end{equation}
where $\psi_{ij}^k$ is the angular separation between event $i$ and source $j$, and $\sigma_i^k$ is the event's angular uncertainty.
The quantity $E_{\mathrm{proxy},i}^k$ is the energy proxy (i.e., the reconstructed muon energy) of the event, $\delta_j$ is the declination of source $j$, and $t_i^k$ is the event arrival time.
For time-integrated analysis,  $S_{ij,\mathrm{temporal}}^k(t_i^k) = 1$.

For event $i$ from detector phase $k$, the reconstructed direction is $(\alpha_i^k,\delta_i^k)$.
The angular separation from a source at $(\alpha_j,\delta_j)$ satisfies
\begin{equation}
 \cos\psi_{ij}^k=\sin\delta_i^k\sin\delta_j+
 \cos\delta_i^k\cos\delta_j\cos(\alpha_i^k-\alpha_j).
\end{equation}
\system adopts a circular two-dimensional Gaussian approximation for angular errors in the local tangent plane, as in earlier IceCube point-source searches and the public-data {\tt SkyLLH} analysis~\cite{IceCube:2010nca,IceCube:2023ihk}.
The two orthogonal error components are independent Gaussians with standard deviation $\sigma_i^k$, so the angular separation $\psi_{ij}^k$ follows a Rayleigh distribution.
Using the spherical annulus area $d\Omega=2\pi\sin\psi_{ij}^k\,d\psi_{ij}^k$ gives the corresponding signal spatial PDF per unit solid angle, as implemented in {\tt SkyLLH}~\cite{skyllhv26code}:
\begin{equation}
 S_{ij,\mathrm{spatial}}^k(\psi_{ij}^k,\sigma_i^k)=
 \frac{\psi_{ij}^k}{2\pi(\sigma_i^k)^2\sin\psi_{ij}^k}
 \exp\!\left[-\frac{(\psi_{ij}^k)^2}{2(\sigma_i^k)^2}\right].
 \label{eq:psf}
\end{equation}
For small separations, $\psi_{ij}^k/\sin\psi_{ij}^k\simeq1$, recovering the usual two-dimensional Gaussian expression. 

The signal energy PDF gives the expected reconstructed-muon-energy distribution for a source with the adopted neutrino spectrum.
Let $\phi_\nu(E_\nu;\boldsymbol{\theta})$ denote the neutrino spectrum, with $\phi_{\nu,j}(E_\nu;\boldsymbol{\theta})\propto\phi_\nu(E_\nu;\boldsymbol{\theta})$.
The weighting factors $w_{j,\mathrm{model}}$ below specify the relative source normalizations.
We can choose any normalization of $\phi_\nu$ because it cancels in the normalized PDFs and signal fractions.
The signal energy PDF combines this shape with the effective area and energy response~\cite{Chang:2022hqj,IceCube:2023ihk} as
\begin{equation}
 S_{ij,\mathrm{energy}}^k(E_{\mathrm{proxy},i}^k\mid\delta_j,\boldsymbol{\theta})
 \propto\int dE_\nu\,
 \phi_\nu(E_\nu;\boldsymbol{\theta})\,
 A_{\mathrm{eff}}^k(E_\nu,\delta_j)\,
 R^k(E_{\mathrm{proxy},i}^k\mid E_\nu,\delta_j).
 \label{eq:signal_energy}
\end{equation}
We obtain $R^k(E_{\mathrm{proxy}}\mid E_\nu,\delta_j)$ from the released smearing matrix~\cite{Abbasi:2026ehs} to describe the energy-proxy distribution for a parent neutrino with true energy $E_\nu$ and source declination $\delta_j$.
We normalize the PDF over the supported energy-proxy range so that $\int d\log_{10}(E_{\mathrm{proxy}}/\mathrm{GeV})\,S_{ij,\mathrm{energy}}^k(E_{\mathrm{proxy}}\mid\delta_j,\boldsymbol{\theta})=1$.
For conventional power laws, \system interpolates pre-calculated tables covering $-6.1\leq-s\leq-0.9$, including margins around the default fit range.
For a spectrum outside that range or a supplied non-power-law shape, it calculates the signal energy PDFs and detector acceptance from the same detector response before the likelihood fitting.
In a stacking analysis, \system evaluates the energy PDF at each source declination and includes it in that source's contribution.

We use the effective areas and smearing matrices released together with the data without applying additional corrections for on-shell $W$-boson production in neutrino--nucleus interactions~\cite{Seckel:1997kk,Alikhanov:2015kla,Zhou:2019vxt,Zhou:2019frk} or for final state radiation in neutrino interactions~\cite{Plestid:2024bva}.
For the current data, we expect the shifts in the fitted fluxes and spectral parameters from these effects to be smaller than their statistical uncertainties.
Accounting for both processes will be important as larger event samples improve the precision of source flux and spectrum measurements.

The signal temporal PDF, used for time-dependent analysis, describes the arrival-time distribution for an adopted neutrino light curve $g_j(t)$.
We set $I_k(t)=1$ during the selected detector uptime intervals and $I_k(t)=0$ otherwise.
\system calculates the signal temporal PDF as
\begin{equation}
 S_{ij,\mathrm{temporal}}^k(t_i^k)=
 \frac{I_k(t_i^k)\,g_j(t_i^k)}
 {\displaystyle\int dt\,I_k(t)\,g_j(t)}.
 \label{eq:signal_temporal}
\end{equation}
\system uses two widely used temporal profiles, Gaussian and Box functions, to describe the neutrino light curve~\cite{IceCube:2011ai,IceCube:2015usw,IceCube:2023ihk}:
\begin{equation}
 g_j(t)=
 \begin{cases}
 \exp\!\left[-\dfrac{(t-t_{0,j})^2}{2\sigma_{t,j}^2}\right],
 & \text{Gaussian},\\[5pt]
 \begin{cases}
 1, & |t-t_{0,j}|\leq\Delta t_j/2,\\[3pt]
 0, & |t-t_{0,j}|>\Delta t_j/2,
 \end{cases}
 & \text{Box},
 \end{cases}
 \label{eq:time_profiles}
\end{equation}
Here $t_{0,j}$ is the center of the neutrino light curve, $\sigma_{t,j}$ is the Gaussian standard deviation, and $\Delta t_j$ is the box duration.

For a time-integrated stacking analysis, the signal PDF is a normalized weighted sum of the individual-source signal PDFs~\cite{IceCube:2010nca,Zhou:2021rhl},
\begin{equation}
 \begin{multlined}
 S_i^k(\boldsymbol{\theta})=\\
 \frac{\displaystyle\sum_j w_{j,\mathrm{model}}w_{j,\mathrm{acc}}^k(\boldsymbol{\theta})
 S_{ij,\mathrm{spatial}}^k(\psi_{ij}^k,\sigma_i^k) S_{ij,\mathrm{energy}}^k(E_{\mathrm{proxy},i}^k\mid\delta_j,\boldsymbol{\theta})}
 {\displaystyle\sum_j w_{j,\mathrm{model}}w_{j,\mathrm{acc}}^k(\boldsymbol{\theta})}.
 \end{multlined}
 \label{eq:stack}
\end{equation}
Here $w_{j,\mathrm{model}}w_{j,\mathrm{acc}}^k(\boldsymbol{\theta})$ is proportional to the expected number of detected signal events from source $j$ in detector phase $k$.
We use $w_{j,\mathrm{model}}$ to specify the relative neutrino fluxes of the sources under the adopted emission hypothesis.
\system requires finite weighting factors satisfying $w_{j,\mathrm{model}}\geq0$ and normalizes them before use.

The factor $w_{j,\mathrm{acc}}^k(\boldsymbol{\theta})$ accounts for detector acceptance, which depends on the source declination and the assumed neutrino spectrum.
For detector livetime $T_k$ and effective area $A_{\mathrm{eff}}^k$, we calculate it from
\begin{equation}
 w_{j,\mathrm{acc}}^k(\boldsymbol{\theta})\ \propto\ T_k
 \int dE_\nu\, A_{\mathrm{eff}}^k(E_\nu,\delta_j)
 \phi_\nu(E_\nu;\boldsymbol{\theta}).
 \label{eq:acceptance}
\end{equation}
\begin{samepage}
The fraction of the total signal expected from detector phase $k$ is
\begin{equation}
 f_k(\boldsymbol{\theta})\equiv
 \frac{\sum_j w_{j,\mathrm{model}}w_{j,\mathrm{acc}}^k(\boldsymbol{\theta})}
 {\sum_{\ell}\sum_j w_{j,\mathrm{model}}w_{j,\mathrm{acc}}^{\ell}(\boldsymbol{\theta})},
 \qquad \sum_k f_k(\boldsymbol{\theta})=1.
 \label{eq:datasetfraction}
\end{equation}
\end{samepage}

For each time-dependent single-source analysis, \system multiplies the steady acceptance in Eq.~\eqref{eq:acceptance} by $\int dt\,I_k(t)g_j(t)/T_k$ and normalizes the resulting yields across detector phases.
This accounts for the overlap of the neutrino temporal profile with detector uptime and preserves the dependence of the signal fractions on the adopted spectral shape.
The reported TS follows Eq.~\eqref{eq:ts}, without an additional duration penalty.

\subsection{Background distributions and PDF tables}
\label{app:pdfs}

The background PDF factorizes into spatial, energy, and temporal PDFs~\cite{Chang:2022hqj,IceCube:2023ihk},
\begin{equation}
 \begin{aligned}
 \mathcal{B}_i^k
 &=B_{i,\mathrm{spatial}}^k\,
 B_{i,\mathrm{energy}}^k(E_{\mathrm{proxy},i}^k\mid\sin\delta_i^k)\,
 B_{i,\mathrm{temporal}}^k(t_i^k)\\
 &=\frac{p_{\mathrm{sp}}^k(\sin\delta_i^k)}{2\pi}\,
 B_{i,\mathrm{energy}}^k(E_{\mathrm{proxy},i}^k\mid\sin\delta_i^k)\,
 B_{i,\mathrm{temporal}}^k(t_i^k).
 \end{aligned}
 \label{eq:background}
\end{equation}
The time-integrated analysis also omits the background temporal PDF factor.
Here $p_{\mathrm{sp}}^k$ is normalized in $\sin\delta$, and the background energy PDF $B_{i,\mathrm{energy}}^k$ is normalized in $\log_{10}(E_{\mathrm{proxy}}/\mathrm{GeV})$ at each declination.
Both signal and background PDFs use the same solid-angle and energy measure, so the likelihood ratio needs no additional declination Jacobian.

For a background event rate that is constant during detector uptime, the temporal PDF is
\begin{equation}
 B_{i,\mathrm{temporal}}^k(t_i^k)=\frac{I_k(t_i^k)}{T_k},
 \qquad T_k\equiv\int dt\,I_k(t).
 \label{eq:background_temporal}
\end{equation}
It is uniform over the detector uptime intervals and zero during gaps, with $\int dt\,B_{i,\mathrm{temporal}}^k(t)=1$.

We calculate the background spatial and energy PDFs from the experimental events~\cite{Zhou:2021rhl,Chang:2022hqj} and store them as tables that \system loads during the analysis.
The background event distribution is uniform in RA.
The full 14-year DR2 analysis uses the v3 tables described in Appendix~\ref{app:background_details}.

When an analysis selects a declination band instead of the whole sky, \system renormalizes the background spatial PDF over that band.
\system saves the event selection and normalization so background trials can repeat the same fit.

\section{Statistical significance and flux constraints}
\label{app:statistics}

This appendix details the significance and flux calculations briefly discussed in Sec.~\ref{sec:data}.
We describe analytic significance estimates (Appendix~\ref{app:analytic}), significance calibration with background trials (Appendix~\ref{app:trials}), and flux upper limits and measurements (Appendix~\ref{app:limits}).

\subsection{Analytic significance estimates and catalog scans}
\label{app:analytic}

\system estimates local significance using a half-$\chi_{m+1}^2$ approximation, where $m$ is the number of fitted spectral parameters and the additional degree of freedom comes from the fitted signal count $n$.
A fixed spectrum has $m=0$ and therefore uses $\chi_1^2$~\cite{Cowan:2010js}, while the default spatial+energy analysis fits $s$, giving $m=1$ and $\chi_2^2$.
The estimated local $p$-value and one-sided Gaussian significance are
\begin{equation}
 \begin{gathered}
 p_{\mathrm{analytic}}\equiv
 \tfrac12\left[1-F_{\chi_{m+1}^2}(\mathrm{TS})\right],\\
 Z_{\mathrm{analytic}}\equiv \mathrm{CDF}^{-1}(1-p_{\mathrm{analytic}}),
 \end{gathered}
 \label{eq:analytic}
\end{equation}
where $F_{\chi_{m+1}^2}$ is the cumulative distribution function of a chi-square distribution with $m+1$ degrees of freedom and $\mathrm{CDF}$ is the standard normal cumulative distribution function.

These approximations provide initial significance estimates~\cite{Davies1977}.
A more precise significance requires background trials that repeat the same fit using randomized event data~\cite{IceCube:2010nca,IceCube:2023ihk}.
In its analysis report to the user, \system states that a more precise significance requires background trials and emphasizes this need when $Z_{\mathrm{analytic}}>2$.

For a catalog scan, \system automatically estimates trial-corrected (i.e., post-trial) $p$-values using the correction adopted in, e.g., Ref.~\cite{Zhou:2021rhl},
\begin{equation}
 p_{\mathrm{post}}=1-\left(1-p_{\mathrm{pre}}\right)^{N_{\mathrm{src}}},
 \label{eq:catalog_trials_correction}
\end{equation}
where $p_{\mathrm{pre}}$ is the pre-trial $p$-value of the source and $N_{\mathrm{src}}$ is the number of sources searched.
It also reports the corresponding one-sided Gaussian significance, $Z_{\mathrm{post}}=\mathrm{CDF}^{-1}(1-p_{\mathrm{post}})$.

\subsection{Background trials with scrambled events}
\label{app:trials}

The analytic $p$-values and significances in Appendix~\ref{app:analytic} are approximations that may not accurately describe the background TS distribution for a particular analysis.
A more rigorous evaluation requires background trials, which repeat the analysis using scrambled events to measure how often background gives $\mathrm{TS}\geq\mathrm{TS}_{\mathrm{obs}}$, 
where $\mathrm{TS}_{\mathrm{obs}}$ is the TS obtained from the observed data.

\system uses the same event-scrambling methods as IceCube and {\tt SkyLLH} analyses~\cite{IceCube:2010nca,IceCube:2011ai,skyllhscrambling}.
For time-integrated analyses, \system draws each event's RA independently and uniformly on $[0,2\pi)$, leaving its other properties unchanged.
Each trial repeats the same analysis with the same scientific settings and fit procedure.
For time-dependent analyses, \system selects a detector run for each event with probability proportional to the number of events recorded in that run, then draws a new arrival time uniformly between the run's start and stop.
Here, a detector run corresponds to a data-taking interval listed in the uptime tables provided with the data release.
Each detector phase contains many such runs~\cite{dr2web}.
\system assigns a new arrival time while keeping each event's zenith and azimuth in the detector's local coordinate system fixed.
Because Earth rotates, the same local direction corresponds to different celestial coordinates at different times.
\system recalculates the event's RA at the new arrival time.
Each trial repeats the original fit, including any search over $t_{0,j}$ and the width ($\sigma_{t,j}$ or $\Delta t_j$) when these are fitted.
The agent saves completed trials so the user can stop and resume the calculation without repeating them.

For $N$ background trials, let $K$ be the number of trials whose TS equals or exceeds the observed value $\mathrm{TS}_{\mathrm{obs}}$.
The empirical $p$-value and one-sided Gaussian significance are
\begin{equation}
 \widehat p\equiv\frac{K+1}{N+1},
 \qquad Z\equiv \mathrm{CDF}^{-1}\!\left(1-\widehat p\right).
 \label{eq:empirical}
\end{equation}
When $K=0$, the finite-trial correction gives $\widehat p=1/(N+1)$, the minimum reported value for $N$ trials~\cite{Phipson:2010}.

\system reports 68\% and 95\% Monte Carlo intervals for the uncertainty from the finite trial count.
At the 95\% confidence level, at least 95\% of repeated trial calculations produce intervals that contain the true background $p$-value at every trial count~\cite{howard}.
It also reports standard binomial confidence intervals for a trial count fixed in advance~\cite{ParticleDataGroup:2024cfk}.
The agent saves completed trials so the user can stop and resume the calculation without repeating them.

\subsection{Upper limits and flux measurements}
\label{app:limits}

\system calculates approximate one-sided 95\% flux upper limits using separate fixed-spectrum fits.
By default, it fits the spectra $E_\nu^{-2}$, $E_\nu^{-2.5}$, and $E_\nu^{-3}$, regardless of the primary TS.
For each spectrum, the fit holds $s$ fixed and maximizes the likelihood over $n\geq0$ to obtain $\widehat n$.
The upper limit $n_{\mathrm{UL}}>\widehat n$ satisfies
\begin{equation}
 2\left[\ln\mathcal{L}(\widehat n;s)
 -\ln\mathcal{L}(n_{\mathrm{UL}};s)\right]
 =\left[\mathrm{CDF}^{-1}(0.95)\right]^2\simeq2.71.
 \label{eq:limit}
\end{equation}
The number $2.71$ follows from the asymptotic likelihood approximation for a one-sided 95\% upper limit~\cite{Cowan:2010js,ParticleDataGroup:2024cfk}.
If Eq.~\eqref{eq:limit} has no solution within the specified $n$ bounds, the agent reports that the upper limit could not be determined.
\system converts $n_{\mathrm{UL}}$ to a flux upper limit using the detector responses.

\system also reports a flux measurement when the fit gives $\mathrm{TS}>9$.
\system obtains the best-fit flux from $\widehat n$ and $\widehat{s}$ using the detector response.
For a power law, the measured $\nu_\mu+\bar\nu_\mu$ flux has the form
\begin{equation}
 \phi_\nu(E_\nu)
 =\phi_{\nu,0}\left(\frac{E_\nu}{1~\mathrm{TeV}}\right)^{-s},
 \label{eq:measured_flux}
\end{equation}
where $\phi_{\nu,0}$ is the differential flux normalization at 1~TeV.
For a stacking analysis, $\phi_\nu(E_\nu)=\sum_j\phi_{\nu,j}(E_\nu)$ is the sum of the source fluxes.

To determine confidence intervals for the flux normalization while accounting for its correlation with the spectral index, \system uses a profile likelihood.
For each trial value of $\phi_{\nu,0}$, it maximizes the likelihood over $s$, converting the flux to $n$ using the $s$-dependent detector acceptance.
The profile-likelihood ratio is
\begin{equation}
 q(\phi_{\nu,0})\equiv2\left[
 \ln\mathcal{L}(\widehat\phi_{\nu,0},\widehat{s})
 -\max_s\ln\mathcal{L}(\phi_{\nu,0},s)
 \right],
 \label{eq:flux_profile}
\end{equation}
where $\widehat\phi_{\nu,0}$ and $\widehat{s}$ are the best-fit values.
The two-sided 68\% and 95\% confidence intervals have endpoints at $q(\phi_{\nu,0})=1$ and $q(\phi_{\nu,0})\simeq3.84$, respectively~\cite{Wilks1938,ParticleDataGroup:2024cfk}.
\system obtains intervals for $s$ by profiling over $n$.
Joint $\phi_{\nu,0}$--$s$ contours satisfy $2[\ln\mathcal{L}(\widehat\phi_{\nu,0},\widehat{s})-\ln\mathcal{L}(\phi_{\nu,0},s)]=2.30$ and $5.99$ for the same confidence levels~\cite{ParticleDataGroup:2024cfk}.
When $s$ is fixed, only $\phi_{\nu,0}$ varies.

The confidence levels of both the upper limits and flux-measurement intervals rely on asymptotic approximations.
A more rigorous evaluation requires signal-injection tests to assess bias in the fitted flux and check the coverage of the confidence intervals.

The confidence levels of both the upper limits and flux-measurement intervals rely on asymptotic approximations.
A more rigorous evaluation requires signal-injection tests to assess bias in the fitted flux and check the coverage of the confidence intervals.

In Sec.~\ref{sec:magnetars}, \system wrote and ran new code to perform signal-injection trials and construct the flux upper limit for the stacking analysis on magnetars, following the method used in the reference analysis.
This demonstrates that \system is well instructed to develop new analysis functions at users' request when the built-in software does not yet support them.

\section{Weighting diagnostics and event selection}
\label{app:analysis_choices}

This appendix details the source weighting diagnostics (Appendix~\ref{app:weighting_choices}) and event selection (Appendix~\ref{app:event_selection}) briefly discussed in Sec.~\ref{sec:before_analysis}.

\subsection{Weighting diagnostics}
\label{app:weighting_choices}

The user's chosen weighting scheme could assign most of the assumed neutrino emission to a few sources, even in a large catalog. 
Before fitting, the agent reports the largest normalized weighting factor ($w_{\mathrm{max},\mathrm{model}}$) and the effective source count ($N_{\mathrm{eff}}^{\mathrm{src}}$) for the selected catalog column so users can assess whether it matches their neutrino emission hypothesis. 
For $N$ sources, let $w_{j,\mathrm{model}}\geq0$ denote the normalized weighting factor of source $j$, with $\sum_{j=1}^{N}w_{j,\mathrm{model}}=1$.
These diagnostics are
\begin{align}
 w_{\mathrm{max},\mathrm{model}}&\equiv\max_j w_{j,\mathrm{model}},
 \label{eq:appendix_maximum_source_weight}\\
 N_{\mathrm{eff}}^{\mathrm{src}}&\equiv\frac{1}{\sum_{j=1}^{N}w_{j,\mathrm{model}}^2}.
 \label{eq:appendix_effective_source_count}
\end{align}
For the normalized weighting factors in Eqs.~\eqref{eq:appendix_maximum_source_weight} and~\eqref{eq:appendix_effective_source_count}, the effective source count satisfies $1\leq N_{\mathrm{eff}}^{\mathrm{src}}\leq N$.
If all $N$ sources have equal weighting factors, each has $w_{j,\mathrm{model}}=1/N$, giving $N_{\mathrm{eff}}^{\mathrm{src}}=N$.
At the other extreme, if one source has $w_{j,\mathrm{model}}=1$ and all other sources have $w_{j,\mathrm{model}}=0$, then $N_{\mathrm{eff}}^{\mathrm{src}}=1$.
More generally, if the total weighting factor is distributed equally among $m$ sources, each of these sources has $w_{j,\mathrm{model}}=1/m$ and all others have $w_{j,\mathrm{model}}=0$, giving $N_{\mathrm{eff}}^{\mathrm{src}}=m$.
Multiplying all unnormalized weighting factors by the same positive constant does not change their relative values or the normalized weighting factors, so both $w_{\mathrm{max},\mathrm{model}}$ and $N_{\mathrm{eff}}^{\mathrm{src}}$ remain unchanged.

For example, for a catalog with $N=10$, we can compare three distributions of weighting factors.
If one source carries 90\% of the total weighting factor and the remaining 10\% is distributed equally among the other nine, then $N_{\mathrm{eff}}^{\mathrm{src}}=[0.9^2+9(0.1/9)^2]^{-1}\simeq1.23$.
The effective source count is close to one because one source carries most of the total weighting factor.
If the total weighting factor is instead distributed equally among all ten sources, each has $w_{j,\mathrm{model}}=0.1$ and the effective source count is $N_{\mathrm{eff}}^{\mathrm{src}}=10$.
If two sources each carry 20\% and the other eight each carry 7.5\%, then $N_{\mathrm{eff}}^{\mathrm{src}}=[2(0.2)^2+8(0.075)^2]^{-1}=8$.
In this case, the weighting factors are more evenly distributed than in the 90\% example, but the two sources with larger weighting factors reduce the effective source count below ten.
On the other hand, a large catalog can also have an effective source count close to one.
For example, if one source carries 99\% of the total weighting factor, $w_{\mathrm{max},\mathrm{model}}=0.99$ and $N_{\mathrm{eff}}^{\mathrm{src}}\simeq1.02$, even for a catalog containing 1,000 sources.
The other 999 sources together carry only 1\% of the total weighting factor.

Users can also request source-contribution checks that repeat the stacking fit after removing individual sources or groups.
These checks measure source influence on the observed fit, including the effects of detector acceptance.

\subsection{Event selection}
\label{app:event_selection}

For single-source analyses and catalog scans, \system selects events within a declination band centered on the source.
This selection significantly reduces computation time and has a negligible effect on the fit results.
The band starts with a half-width of $20^\circ$.
If it contains fewer than 20\% of the events from any detector phase, the agent expands it until the selected events account for at least 20\% of the events from each phase.
For a catalog scan, \system applies this selection at each source position.
Stacking analyses use all events unless the user requests a cut.
Users can also request all events for a single-source analysis or specify explicit declination bounds.

\section{Behind the scenes: analysis execution and records}
\label{app:interface}

This appendix describes how \system executes analysis commands and saves results behind the scenes.
We provide these details for curious readers, but users do not need them to use \system.
We describe the commands that the agent generates and executes on the user's behalf (Appendix~\ref{app:running_analyses}) and the saved results and calculation records (Appendix~\ref{app:saved_results}).

\subsection{How \system runs analyses}
\label{app:running_analyses}

After determining the source information and scientific choices described in Sec.~\ref{sec:before_analysis}, \system generates and executes commands for \path{run_agent_wrap.py}.
Listings~\ref{lst:single_source} and~\ref{lst:catalog} show a single-source example and a catalog scan with stacking analysis, respectively.
The agent runs these commands on the user's behalf, so users can work through conversation without learning the command syntax below.
The \code{--request} argument saves the user's original words, while the remaining arguments specify the source information and analysis settings determined from that conversation.
The \code{create-single} and \code{create-catalog} commands check the inputs, create an analysis task, and return its identifier and status.
The agent executes the \code{run} command with that task identifier only after the status is \status{READY}, indicating that the required choices and input checks are complete.

\noindent\begin{minipage}{\linewidth}
\begin{lstlisting}[style=agentcode,label={lst:single_source},caption={
\system creates a single-source analysis from its natural-language interaction with the user, checks the inputs, and runs the analysis.
}]
.venv/bin/python run_agent_wrap.py create-single \
  --request "Analyze TXS 0506+056 and fit n and the power-law exponent" \
  --name "TXS 0506+056" --ra-deg 77.3582 --dec-deg 5.6931

# Execute only after the returned task status is READY:
.venv/bin/python run_agent_wrap.py run --task <task_id> --workers 2
\end{lstlisting}
\end{minipage}

\noindent\begin{minipage}{\linewidth}
\begin{lstlisting}[style=agentcode,label={lst:catalog},caption={
\system creates a catalog scan and stacking analysis with uniform weighting from its natural-language interaction with the user, checks the inputs, and runs both analyses.
}]
.venv/bin/python run_agent_wrap.py create-catalog \
  --request "Single-source and stacking analyses with equal weighting" \
  --catalog catalog.csv --name-column source_name \
  --ra-column ra_deg --dec-column dec_deg \
  --catalog-operation scan_and_stack --weight-mode uniform

# Execute only after the returned task status is READY:
.venv/bin/python run_agent_wrap.py run --task <task_id> --workers 2
\end{lstlisting}
\end{minipage}

The \code{--workers N} option tells the \code{run} command how many CPU threads to use, with \code{--workers 1} as the default.
For a catalog scan, the agent can use these resources to fit up to $N$ sources at the same time.
For a single-source or stacking analysis, the option allows up to $N$ CPU threads within one fit.
For example, \code{--workers 2} requests two CPU threads for the single-source example in Listing~\ref{lst:single_source}.
For background trials, the separate \code{dr2-empirical-scramble} command uses two CPU threads by default and runs two trial fits at the same time.
The requested number of CPU threads changes the computing resources, while the scientific analysis settings stay the same.

The agent uses command arguments to specify the adopted neutrino spectrum.
For the power law in Eq.~\eqref{eq:spectrum}, the interface takes the signed exponent $-s$, so a falling spectrum with $s=2.5$ uses \code{--gamma -2.5}.
To keep this spectrum fixed during the fit, the agent also supplies \code{--gamma-mode fixed}.
By comparison, \code{--gamma +2.5} specifies the rising spectrum $E_\nu^{+2.5}$.
For a tabulated spectrum, \code{--spectrum-file} identifies the input file and \code{--spectrum-energy-unit} specifies its energy unit.
The \code{--spectrum-kind} argument states whether the tabulated values represent $dN/dE_\nu$, $E_\nu\,dN/dE_\nu$, or $E_\nu^2\,dN/dE_\nu$.
The \code{--spectrum-outside-support} argument specifies whether the spectrum is zero outside the tabulated energy range or follows a power-law extrapolation.

In the default power-law fit, both $n$ and the spectral index vary.
The allowed ranges are $0\leq n\leq10^4$ and $-6\leq-s\leq-1$, and the fit starts at $n=10$ and $-s=-3$.
If the user supplies $s$, the agent fixes it unless the request asks for a fit or gives that value as an initial value.

If a calculation fails because of a software or file-access problem, \system may rerun it once using the same analysis configuration.
The saved records allow users to compare the settings used in the two attempts.
If a best-fit parameter reaches the boundary of its allowed range, the agent asks the user to assess the result and does not widen the range on its own.
See Appendix~\ref{app:saved_results} for more details about these records.

\subsection{How \system saves results and calculation records}
\label{app:saved_results}

For every analysis, \system saves the user's request, the adopted configuration, and how each scientific choice was made.
The record distinguishes choices specified by the user, choices determined from the request or supplied information, and documented defaults.
It groups each attempt at the fit with this record, so users can check whether a repeated calculation used the same inputs and scientific settings.
The saved status records the progress and outcome of the analysis.
\system labels completed results according to whether they pass all checks or have warnings.
The agent also saves a checksum of the request so users can detect changes between attempts.

The output records connect the fitted quantities and result checks to the exact inputs and code version used in the calculation.
\system saves the result checks in JavaScript Object Notation (JSON), a text format that organizes the recorded quantities and their values.
Catalog scans also save comma-separated values (CSV) tables in the original catalog order, together with the checks for each source.
Keeping that order allows users to compare a source's result with its entry in the input catalog.
Users can inspect these records without running the fit again.

For background trials, \system saves the observed reference fit, the empirical $p$-value and corresponding one-sided Gaussian significance calculated from the completed trials, and the Monte Carlo uncertainty intervals for the $p$-value.
It also saves the random seeds and trial identifiers, which allow users to reproduce completed trials and identify which trials have already entered a combined result.
These records allow completed trials to be combined without counting the same trial twice.
Users can also inspect the TS distribution to see how the observed TS compares with those from the background trials.

\section{Validation tests}
\label{app:validation}

This appendix describes the main checks we use to verify that \system carries out the intended analysis and reports results consistent with the saved calculations, as summarized in Sec.~\ref{sec:validation}.
We present the software and upper-limit tests (Appendix~\ref{app:validation_methods}) and the natural-language conversation tests (Appendix~\ref{app:language}).

\subsection{Validation methods}
\label{app:validation_methods}

We use software tests to check that \system runs the analysis with the selected source position and spectrum, handles invalid inputs and failed calculations, and reports the saved fit results correctly.
For the input checks, we submit requests with missing RA or declination, or coordinates outside their allowed ranges.
We check that the agent waits for the user to supply or correct the coordinates before starting the fit.
For completed calculations, we check that the fit used the selected settings and that the analysis report agrees with the saved fitted quantities.
We also check that the agent reports a fit that does not converge as a failure and that it resumes an interrupted calculation with the same scientific settings.
These software tests have passed.

We test whether the upper-limit calculation in Appendix~\ref{app:limits} finds the signal count corresponding to the chosen confidence level.
We determine reference 95\% limits for Gaussian and Poisson toy likelihoods without using the \system upper-limit calculation, then compare these limits with its values of $n_{\mathrm{UL}}$.
We also check that the calculation reports when no solution exists within the allowed $n$ range, rather than returning a bound as a valid upper limit.
These tests passed.

\subsection{Selected natural-language conversations}
\label{app:language}

We use natural-language conversations to check that the AI agent turns the user's words into the intended analysis, asks the user to clarify ambiguous physical choices, and explains the results correctly.
The requests cover all analysis types, and the conversations explicitly invoke the analysis instructions.
The tests examine whether the agent obtains the information needed for the requested analysis before creating or running it.

For example, some requests define the spectral index $s$ through $E_\nu^{-s}$, while others give the signed exponent $-s$ directly.
We check that the agent distinguishes these conventions and asks the user which convention is intended if the request does not define it.
In another example, we provide a catalog containing both radio and X-ray flux columns and request only ``flux weighting''.
Either column would give a different weighting scheme, so we check that the agent asks the user which column to use before creating or running the analysis.

We also test a request for a spectrum that the code could not evaluate at the time of testing.
We check that the agent explains the limitation and keeps the requested spectrum, rather than substituting another spectrum to complete the calculation.
These conversation tests passed.

\section{Construction and comparison of the background PDFs}
\label{app:background_details}

We provides two versions of background PDFs in \system.
Version 1 (v1) follows the algorithm used in the IceCube {\tt SkyLLH} code~\cite{IceCube:2021mzg,IceCube:2023ihk,skyllhv26code}, while version 3 (v3) uses adaptive sliding windows and additional corrections to improve the background energy distributions.
We use v3 by default because it reduces the background energy residuals found with v1 and the resulting spurious signal contributions in stacking analyses.
Appendix~\ref{app:background_construction} describes the construction of both versions, while Appendix~\ref{app:background_residuals} compares their predicted event counts and the stacking TS values obtained for the same random catalogs.

\subsection{Construction of the v1 and v3 background PDFs}
\label{app:background_construction}

We first constructed the v1 background PDFs using the algorithm in the IceCube {\tt\bfseries SkyLLH} code~\cite{IceCube:2021mzg,IceCube:2023ihk,skyllhv26code}.
For each detector phase, {\tt SkyLLH} constructs the background spatial and energy PDFs from the same observed events collected during that detector phase.
Table~\ref{tab:v1_binning} gives the histogram boundaries and bin counts.

For the spatial PDF, {\tt SkyLLH} histograms the events in $\sin\delta$.
The code divides each bin count by the total number of events in that detector phase and the bin width in $\sin\delta$ to obtain a probability density per unit $\sin\delta$.
It interpolates the logarithm of this density at the bin centers with a quadratic spline, then exponentiates the result.
The spatial PDF includes a factor $1/(2\pi)$ for the uniform background distribution in RA, as in Eq.~\eqref{eq:background}.

For the energy PDF, {\tt SkyLLH} fills a two-dimensional histogram in $\log_{10}(E_{\mathrm{proxy}}/\mathrm{GeV})$ and $\sin\delta$.
Within each declination bin, the code divides each energy-bin count by the number of events in that declination bin and by the log-energy bin width.
It then smooths the energy distribution within each declination bin by averaging the density in each energy bin with the densities in its immediate neighbors, using equal weights.
At the lower and upper energy boundaries, it averages the density in the boundary bin with that in its one available neighbor.
Before taking the logarithm, the code replaces zero-valued bins with $10^{-3}$ times the smallest positive density in the smoothed histogram for that detector phase.
It interpolates the logarithm of the smoothed histogram with a cubic spline along the log-energy axis.
Along $\sin\delta$, the interpolation keeps the energy density nearly constant within each declination bin, with narrow transitions across the declination-bin boundaries.

\begin{table}[htbp]
\caption{
Histogram binning used for the v1 background PDFs.
We take these settings from the IceTracks-DR2 dataset definitions in {\tt SkyLLH}~\cite{skyllhv26code}.
The column labeled ``Bins per interval'' gives the number of equal-width bins in each listed $\sin\delta$ interval.
All energy histograms have bin width $\Delta\log_{10}(E_{\mathrm{proxy}}/\mathrm{GeV})=0.125$.
}
\label{tab:v1_binning}
\setlength{\tabcolsep}{8pt}
\begin{ruledtabular}
\begin{tabular}{cccc}
Detector phase & $\sin\delta$ intervals & Bins per interval & $\log_{10}(E_{\mathrm{proxy}}/\mathrm{GeV})$ range \\
\hline
IC40 & $[-1,-0.25]$, $[-0.25,0]$, $[0,1]$ & $10,10,10$ & $[2,9]$ \\
IC59 & $[-1,-0.95]$, $[-0.95,-0.25]$, $[-0.25,0.05]$, $[0.05,1]$ & $2,25,15,10$ & $[2,9.5]$ \\
IC79 & $[-1,1]$ & $49$ & $[1,9]$ \\
IC86 I--XI & $[-1,-0.93]$, $[-0.93,-0.3]$, $[-0.3,0.05]$, $[0.05,1]$ & $4,10,9,18$ & $[1,9.5]$ \\
\end{tabular}
\end{ruledtabular}
\end{table}

We found that v1 underestimates the observed event counts in some energy-proxy bins, which can increase the stacking TS (Appendix~\ref{app:background_residuals}).
To reduce the background energy residuals, we developed v3 using an adaptive sliding-window algorithm to estimate the background spatial and energy PDFs at each declination.
We use the same factorization and normalization conventions as in v1 (Eq.~\eqref{eq:background}).
We use the parameters $k_{\mathrm{sp}}$, $k_{\delta}$, and $k_E$ to specify the target event counts for the spatial window, the declination window for the energy PDF, and the energy-smoothing window.
For the spatial PDF, we select a window in $\sin\delta$ containing approximately $k_{\mathrm{sp}}$ events around each declination.

For the energy PDF, we select approximately $k_{\delta}$ events with nearby $\sin\delta$ values.
We spread each event's contribution uniformly over a log-energy interval containing approximately $k_E$ local events and average these contributions to estimate the energy distribution.
For fixed target counts, the windows are narrower where events are more densely distributed in $\sin\delta$ or log-energy.

To avoid zero background density, we mix the estimated energy density with a component uniform over the PDF's log-energy range.
Its probability weight is $1/(N_{\mathrm{local}}+1)$, where $N_{\mathrm{local}}$ is the event count in the declination window.
The density estimated from these events carries the remaining probability weight.

We choose the smoothing parameters for each detector phase by comparing how well candidate PDFs predict events excluded from their construction.
We divide the events into groups and calculate the mean negative log-likelihood with each group excluded in turn.
Among candidates within one standard error of the minimum, we choose the largest smoothing scale to reduce fluctuations from sparse event counts.
We do not use source-fit results to select these parameters.

For IC40, IC59, and IC79, we apply an additional correction to reduce broad differences between observed and PDF-predicted energy counts.
In a centered window of width $0.4$ in $\log_{10}(E_{\mathrm{proxy}}/\mathrm{GeV})$, let $N_{\Delta E}$ and $M_{\Delta E}$ denote the observed and predicted counts for the selected local events, with the prediction excluding the uniform component.
We multiply the estimated density by $(N_{\Delta E}+c)/(M_{\Delta E}+c)$, where $c$ suppresses corrections for small counts.
We renormalize the corrected density before adding the same uniform component.
We choose the correction parameters using the same excluded-event criterion.
For IC86, we use the adaptive estimate without this correction.
Table~\ref{tab:background_pdf_settings} gives the parameters used for v3.

\begin{table}[t]
\caption{
Parameters used to construct the v3 background PDFs.
The target counts $k_{\mathrm{sp}}$, $k_{\delta}$, and $k_E$ set the widths of the spatial window, the declination window used for the energy PDF, and the energy-smoothing window, respectively.
Events with identical $\sin\delta$ or energy-proxy values can make a window contain more events than its target count.
The correction width is measured in $\log_{10}(E_{\mathrm{proxy}}/\mathrm{GeV})$.
A dash denotes no additional energy-shape correction.
}
\label{tab:background_pdf_settings}
\setlength{\tabcolsep}{8pt}
\begin{ruledtabular}
\begin{tabular}{cccccc}
Detector phase & $k_{\mathrm{sp}}$ & $k_{\delta}$ & $k_E$ & Correction width & $c$ \\
\hline
IC40 & 1126 & 563 & 40 & 0.4 & 10 \\
IC59 & 1320 & 660 & 45 & 0.4 & 1 \\
IC79 & 1270 & 2539 & 66 & 0.4 & 1 \\
IC86 I--XI & 5154 & 20617 & 22 & --- & --- \\
\end{tabular}
\end{ruledtabular}
\end{table}

\subsection{v1 versus v3: energy residuals and spurious signal contributions}
\label{app:background_residuals}

\begin{figure}[t]
\centering
\includegraphics[width=0.94\textwidth]{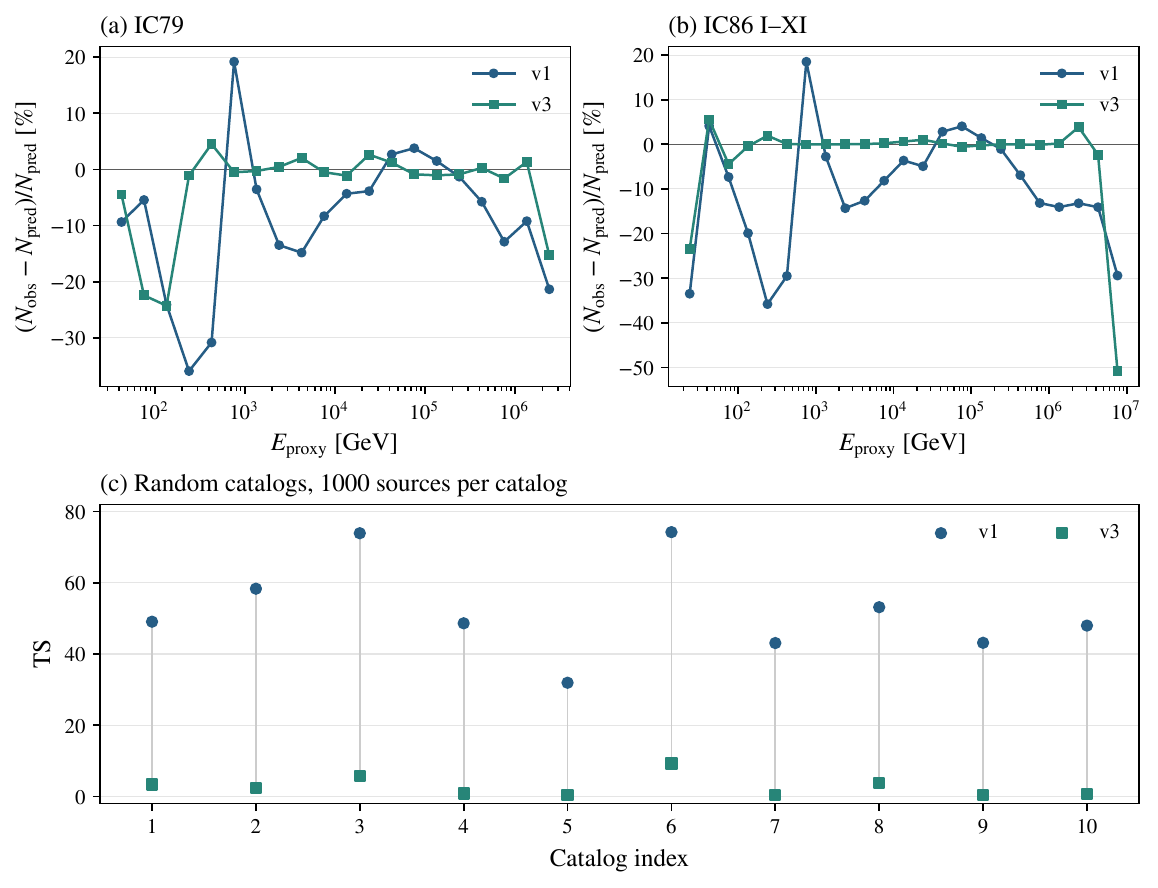}
\caption{
Comparison of the v1 and v3 background energy PDFs and their effect on stacking analyses.
Panels (a) and (b) show the differences between the observed and predicted event counts in energy-proxy bins for IC79 and the combined IC86 I--XI data.
We express these differences as $100(N_{\mathrm{obs}}-N_{\mathrm{pred}})/N_{\mathrm{pred}}$, where $N_{\mathrm{obs}}$ and $N_{\mathrm{pred}}$ are the observed and predicted counts.
To calculate $N_{\mathrm{pred}}$, we integrate the background energy PDF over each energy-proxy bin at each event's declination and sum these probabilities over all events from that detector phase.
We show only bins with at least 20 observed events in these panels, while using the full event selection for the analyses.
Panel (c) shows the stacking TS for random catalogs of 1,000 sources drawn from an isotropic sky distribution, with equal weighting factors and fitted $n$ and $s$.
For each random catalog, the v1 and v3 fits use the same source positions and analysis settings except for the background PDFs.
All random catalogs use the same IceTracks-DR2 events.
The v1 background mismatches produce spurious signal contributions, while v3 significantly reduces the residuals and, as a result, the TS values.
}
\label{fig:background_pdf_comparison}
\end{figure}

Fig.~\ref{fig:background_pdf_comparison} shows the differences between the observed event counts and the counts predicted by the v1 and v3 background energy PDFs.
Panels (a) and (b) show the results for IC79 and the combined IC86 I–XI data, respectively.
For each observed event, we integrate the background energy PDF for its detector phase over each energy-proxy bin at that event’s declination.
We sum these probabilities over all events from that detector phase to obtain the predicted count in each energy-proxy bin.

As shown in the figure, the v1 PDFs, computed using the {\tt SkyLLH} algorithm, predict too many events in some energy bins and too few in others.
For example, they underestimate the observed counts around $E_{\mathrm{proxy}}\simeq0.6$--$1$~TeV.
Where v1 underestimates the background energy density, the stacking fit can increase the likelihood by assigning some of the observed events to a signal contribution.
Especially for large catalogs, background mismatches near many source directions can accumulate in the likelihood and produce a large TS (see panel (c)).
The v3 PDFs reduce these broad residuals, although residuals remain at the lowest and highest energies, where the small number of observed events makes the background energy PDFs difficult to estimate.

We examine the effect of these residuals on the likelihood analysis using stacking analyses of random catalogs of 1000 sources drawn from an isotropic sky distribution.
For these random catalogs, we expect TS values close to zero, apart from statistical fluctuations.
We use all IceTracks-DR2 events and equal weighting, and fit $n\geq0$ and $1\leq s\leq6$.
For each random catalog, the v1 and v3 fits use the same analysis settings except for the background PDFs.
Panel (c) of Fig.~\ref{fig:background_pdf_comparison} shows the large v1 TS values produced by the spurious signal contribution arising from residuals in the background energy PDFs.
The v3 PDFs significantly reduce the energy residuals and, as a result, the TS values for the same random catalogs.

The residuals discussed above arise in the v1 background energy PDFs that we constructed for IceTracks-DR2 using the public {\tt SkyLLH} algorithm.
The published results from IceCube Collaboration uses its internal analysis tools separately from the public {\tt SkyLLH} benchmark.
Background trials can largely remove the effect of these residuals on the significance, provided that the scrambled data represent the background in the absence of a signal.
However, running many trials is expensive because each trial requires a new fit.
A more accurate background energy PDF can make analytic significance estimates more reliable and reduce the need for trials in preliminary analyses.
Flux measurements and upper limits still need an accurate background energy PDF.
If the PDF predicts too few background events, the fit can count some background events as signal and shift the measured flux or its upper limit.

For these reasons, \system uses the v3 background PDFs by default.
The examples in Sec.~\ref{sec:examples} also show that, for both the significance and flux constraints, v3 better reproduces the reference results for NGC~1068 and magnetars than v1.
The v3 background PDFs are not perfect.
They still have residuals at the lowest and highest energies, where the small number of events makes it difficult to construct reliable background PDFs.
We will continue to improve them.

\section{Examples of time-dependent analyses}
\label{app:time-dependent-examples}

\system can also perform time-dependent analysis, including single source-analysis, catalog scan, and stacking analysis. 
The implemented temporal profiles are a Gaussian or a Box signal-time PDF (see Appendix~\ref{app:signal_model} for more details). 
Users need to provide the coordinates (or source names) and time for each source. 
Users will then be asked to specify the time PDF and (1) the source time $t_0$ and the Gaussian standard deviation $\sigma_t$ for Gaussian, or (2) the source time $t_0$ and the start/end time of the box window $(t_{\rm start}, t_{\rm end})$ for Box. When performing stacking analysis, users can define a customized weighting scheme for the catalog (e.g., uniform, proportional to a specific source property like fluence, etc.).

We present a single-source analysis of blazar TXS~0506+056 in Appendix~\ref{app:txs0506+056} and a catalog scan and stacking analysis of the infrared flares associated with dust echoes in Appendix~\ref{app:flaires_td}.

\subsection{Time-dependent single-source analysis: TXS 0506+056}
\label{app:txs0506+056}

\begin{figure*}[!t]
\centering
\begingroup
\makeatletter
\def\@captype{table}
\makeatother
\caption{
The time-dependent single-source analysis results of \system using the IceTracks-DR2 sample for TXS~0506+056~\cite{IceCube:2018cha}.
The results assume a Gaussian time PDF with IceCube's best-fit flare center and duration, and they are obtained using IC86-III, IC86-IV, and IC86-V phases of IceTracks-DR2. 
Our local significance comes from 100,000 matched fixed-window background trials.
}
\label{tab:txs0506+056_td}
\setlength{\tabcolsep}{5pt}
\begin{tabular*}{\textwidth}{@{\extracolsep{\fill}}ccccc}
\hline\hline
Analysis & $\widehat n$ & $\widehat{s}$ & TS & Local significance \\
\hline
\system v3 & 9.4 & 2.1 & 17.3 & $3.9\sigma$ \\
\hline\hline
\end{tabular*}
\endgroup
\par\vspace{1em}
\includegraphics[width=0.5\textwidth]{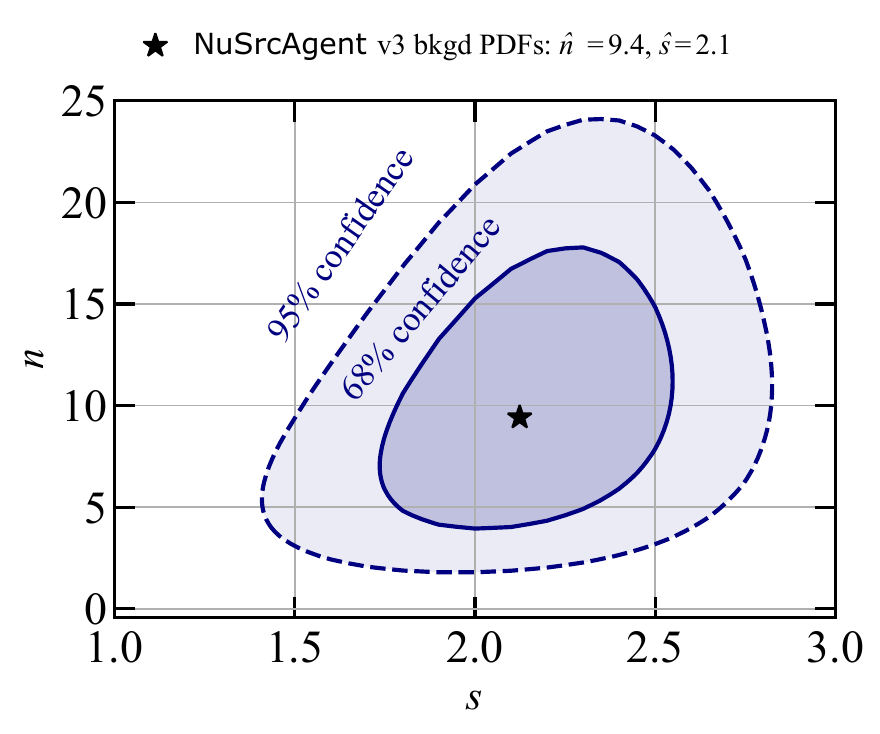}
\par\nointerlineskip
\setlength{\abovecaptionskip}{3pt}
\caption{
The $(n,s)$ likelihood contours reported by \system using the 14-year IceTracks-DR2 sample for TXS~0506+056. Our 68\% and 95\% labels denote the conventional two-parameter $\Delta\mathrm{TS}=2.30$ and $5.99$ levels.
}
\label{fig:txs0506+056_td}
\end{figure*}

Table~\ref{tab:txs0506+056_td} and Fig.~\ref{fig:txs0506+056_td} show the time-dependent analysis results for TXS~0506+056 reported by \system.
After the association of IceCube-170922A with a gamma-ray flare from this blazar, IceCube searched its earlier observations and reported an excess of neutrino events from its direction between September 2014 and March 2015~\cite{IceCube:2018cha}.
Its Gaussian time-profile fit gave a center of MJD~57004 and a duration $T_{\rm W}=110$~days, where $T_{\rm W}=2\sigma_t$.
IceCube reported $3.5\sigma$ evidence after accounting for the searched detector periods and the choice of Gaussian or box time profile~\cite{IceCube:2018cha}.

We fix the Gaussian center at MJD~57004 and its standard deviation at $\sigma_t=55$~days, adopting the best-fit time window reported by IceCube.
We use the public IceTracks-DR2 events from the overlapping IC86-III, IC86-IV, and IC86-V detector phases, the default v3 background PDFs, and a background time PDF uniform over detector uptime.
We fit $n$ and $s$ for a power-law spectrum and run 100,000 matched background trials that repeat this fixed-window fit.
Three trials have TS at least as large as the observed value, giving an add-one local $p$-value of $4.00\times10^{-5}$ ($3.9\sigma$); its 95\% fixed-trial-count Monte Carlo interval is $[6.19\times10^{-6},\,8.77\times10^{-5}]$.
The resulting $\widehat n$ and $\widehat{s}$ agree with IceCube's reported $13\pm5$ and $2.1\pm0.2$ within their quoted uncertainties (Table~\ref{tab:txs0506+056_td}).
Our local significance applies to the specified window only; it does not include IceCube's search over flare centers, durations, detector periods, and time-profile shapes.

Users can obtain the results shown in Table~\ref{tab:txs0506+056_td} and Fig.~\ref{fig:txs0506+056_td} with the following prompts.

\setlength{\fboxsep}{2pt}
\setlength{\leftmargini}{1.3em}
\begin{enumerate}
\addtolength{\linewidth}{-8pt}
\item \promptbox{Do a time-dependent analysis for TXS 0506+056. Use a Gaussian PDF with center = MJD 57004.0 and standard deviation = 55 days. Fit $s$.}
\textbf{Explanation:} \system identifies the source coordinates and uses the three IceTracks-DR2 detector phases overlapping the specified Gaussian window.
It records the fixed window and uniform-livetime background assumption, then fits $n$ and $s$ using v3 background PDFs.
The final response reports the best-fit parameters and TS in Table~\ref{tab:txs0506+056_td}, shows the $(n,s)$ likelihood-drop contours (Fig.~\ref{fig:txs0506+056_td}), and explains that the observed fit alone gives no calibrated significance.
\end{enumerate}

\subsection{Catalog scan and stacking analysis: dust-echo-like infrared flares}
\label{app:flaires_td}

\begin{figure*}[!t]
\centering
\begingroup
\makeatletter
\def\@captype{table}
\makeatother
\caption{Results from \system v3 for the 387 selected sources
from the Flaires catalog~\cite{Necker:2024vfs}.
The source rows are ranked by observed TS.
The local and stacking significances are one-sided Gaussian
equivalents of empirical add-one p-values obtained from
100,000 matched background trials.
Global source significances use the empirical distribution of
the maximum single-source TS across the 387-source catalog.
No catalog-scan correction is applied to the separate stacking hypothesis.}
\label{tab:flaires_td}
\setlength{\tabcolsep}{5pt}
\begin{tabular*}{\textwidth}{@{\extracolsep{\fill}}cccccccc}
\hline\hline
Top sources & Dec [deg] & RA [deg] & IR peak [MJD] &
$\widehat n$ & $\widehat{s}$ & TS &
Local/Global significance \\
\hline
J163334.94+654445.6 & 65.74599 & 248.39556 & 57901.00 &
18.3 & 3.4 & 14.9 & $3.9\sigma/2.1\sigma$ \\
J151813.97-010737.5 & $-1.12709$ & 229.55824 & 58418.79 &
14.1 & 3.7 & 9.8 & $2.9\sigma$/No significance \\
J051007.17+224013.2 & 22.67042 & 77.52970 & 58163.99 &
10.4 & 2.6 & 8.9 & $2.8\sigma$/No significance \\
\hline
Fluence-weighted stacking & \multicolumn{3}{c}{---} &
17.2 & 2.2 & 2.2 & $1.2\sigma/1.2\sigma$ \\
\hline\hline
\end{tabular*}
\endgroup
\par\vspace{1em}
\includegraphics[width=\textwidth]{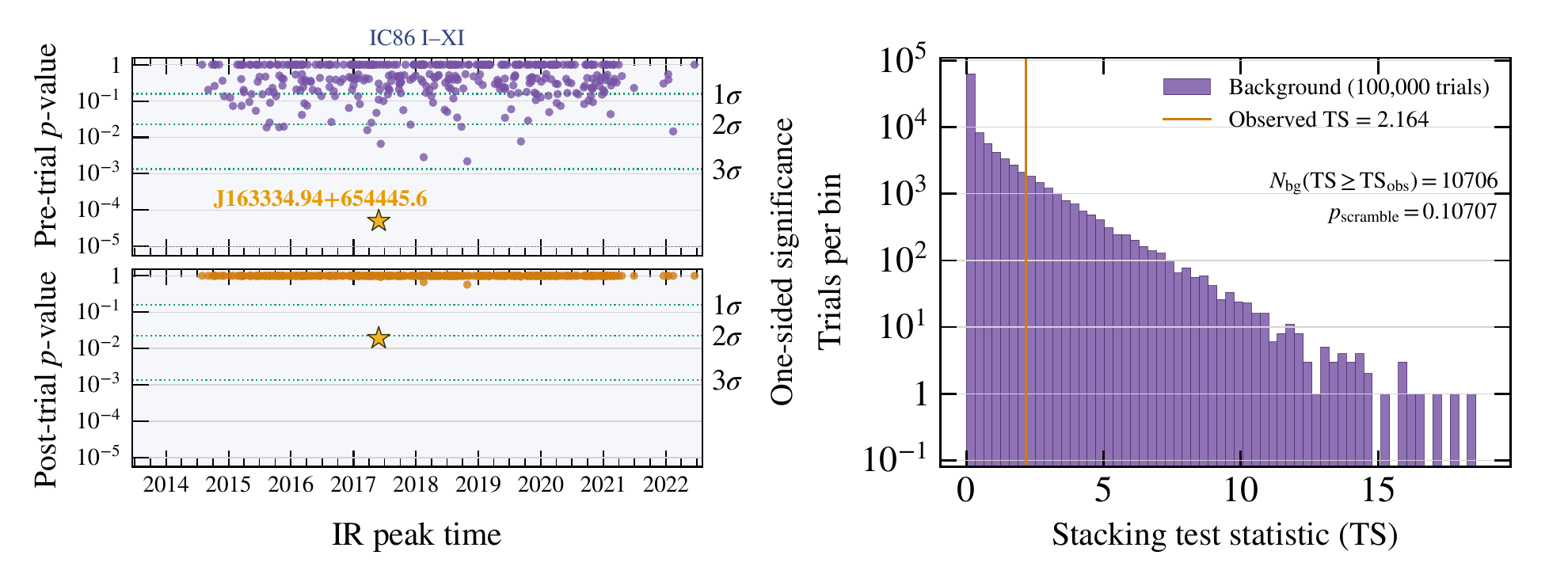}
\par\nointerlineskip
\setlength{\abovecaptionskip}{3pt}
\caption{Results of the 387-source scan and the fluence-weighted stacking analysis using 100,000 background trials.
The left panels show the local and global significance of all sources included in the analysis.
The right panel compares the observed stacking TS with its background-trial distribution.
}
\label{fig:flaires_td}
\end{figure*}

We show an example of using \system to perform a Box time-dependent catalog scan and stacking analysis for the Flaires catalog of infrared flares associated with dust echoes~\cite{Necker:2024vfs}. In Ref.~\cite{Abbasi:2026som}, the IceCube Collaboration selected 393 Northern-sky sources from the catalog and used a one-year Box window to search for high-energy neutrino emission from transient accretion activity. IceCube reported no significance in the stacking analysis with its 10-year track sample.

Table~\ref{tab:flaires_td} shows the leading single-source fits and the fluence-weighted stacking result, while Fig.~\ref{fig:flaires_td} shows the source-search probabilities of the catalog scan and the stacking background-TS distribution.
Following Ref~\cite{Abbasi:2026som}'s selection rules, \system obtained 387 sources from the supplied 823-source catalog.
For each selected source, we fix a box signal-time PDF from 365 days before its infrared peak time, and fit a common spectral index $s$.
We use the public IceTracks-DR2 IC86 I--XI, the default v3 background PDFs, and a background time PDF uniform over detector uptime.
In addition to the fluence-weighted stack, \system performs an independent fit for each catalog source.
The results in Table~\ref{tab:flaires_td} use 100,000 shared seasonal background trials to estimate source-local, catalog-wide, and stacking probabilities.
The leading source has a pre-trial $p$-value of $5\times10^{-5}$ (a local significance of $3.9\sigma$). With 387 sources, this corresponds to a post-trial $p$-value of $0.019$ (a global significance of $2.1\sigma$). For the stacking analysis, the observed TS is $2.2$, and the empirical probability is $0.107$ ($1.2\sigma$).
Note that these results are not a direct reproduction of IceCube's stacked $\mathrm{TS}=0.0$ in Ref.~\cite{Abbasi:2026som} as the selected sources, neutrino event samples, detector responses, and background models are different.

Users can obtain the results shown in Table~\ref{tab:flaires_td} and Fig.~\ref{fig:flaires_td} with the following prompts.

\setlength{\fboxsep}{2pt}
\setlength{\leftmargini}{1.3em}
\begin{enumerate}
\addtolength{\linewidth}{-8pt}
\item \promptbox{Do a time-dependent analysis for the Flaires catalog in [catalog CSV path]. Downsample the sources following the selection rules of the paper 2610.20163.
Use a Box PDF with t\_start = t\_0-365 days, t\_end = t\_0, and t\_0=ir\_peak\_mjd in MJD. For the stacking analysis, the weight of each source should be proportional to the bolometric fluence. Fit $s$.}
\textbf{Explanation:} \system applies the cuts of Ref~\cite{Abbasi:2026som} to the supplied catalog, retaining 387 sources with measured bolometric fluence. The agent then assigns fixed box windows to each source, uses the public IceTracks-DR2 detector phases, fits $n$ and $s$ for each source, and performs stacking analysis with bolometric-fluence weights. The agent will report the results shown in Table~\ref{tab:flaires_td} and produce the $(n, s)$ likelihood contour plots (like the left panel of Fig.~\ref{fig:txs0506+056_td}) from the analysis of leading sources and the stacking analysis.

\par\nobreak\smallskip
\textbf{Agent question:}\enspace
Would you like matched shared seasonal background trials for the source scan and stack? If so, what fixed trial count should I run?
\item \promptbox{Continue with 1e5 background trials.}
\textbf{Explanation:} \system runs 100,000 background-trial fits, where in each trial the agent scrambles the arrival time and RA of neutrino events in all detector phases (see Appendix~\ref{app:trials}) for the same sources, windows, and analysis hypotheses. The agent then reports the significance results in Table~\ref{tab:flaires_td} and Fig.~\ref{fig:flaires_td}. 
As described in Appendix~\ref{app:running_analyses}, \system allows users to specify the number of CPU workers so that the analysis workflow can be accelerated with parallel computation and a cached engine that stores the information that can be shared across different trials. This is particularly useful for running a large number of background trials for catalogs with many sources. If users are running \system on high-performance computing systems with CPU nodes, they can also directly run the analysis through their computational resources by adding  
\fcolorbox{black!35}{black!14}{
    \begin{varwidth}[t]{\dimexpr\linewidth-\promptlabelwidth-2\fboxsep-2\fboxrule-4pt\relax}%
      \small\strut\code{Use CPU nodes from [your own cluster] with [any computational job information]}%
    \end{varwidth}}
to the prompt above.

\end{enumerate}

\twocolumngrid
\end{CJK*}
\bibliographystyle{apsrev4-1}
\AddToHook{env/thebibliography/begin}{\endlinechar=32\relax}
\bibliography{references}
\end{document}